\documentclass[twoside,fleqn]{article}
\usepackage{espcrc2,mathbbol}
\newcommand{\eeq}{\end{equation}}
\newcommand{\beq}{\begin{equation}}
\newcommand{\beqa}{\begin{eqnarray}}
\newcommand{\eeqa}{\end{eqnarray}}
\newcommand{\bea}{\begin{array}}
\newcommand{\eea}{\end{array}}
\newcommand{\Pexp}{{\rm Pexp}}
\newcommand{\diag}{{\rm diag}}
\newcommand{\cO}{{\cal{O}}}

\newcommand{\cA}{{\cal{A}}}
\newcommand{\cD}{{\cal{D}}}
\newcommand{\cF}{{\cal{F}}}
\newcommand{\cV}{{\cal{V}}}
\newcommand{\cS}{{\cal{S}}}
\newcommand{\ff}{{\rm{ff}}}
\newcommand{\Tr}{{\rm Tr}}
\newcommand{\tr}{{\rm tr}}
\newcommand{\veps}{\varepsilon}
\newcommand{\ddz}{{\frac{d}{dz}}}
\newcommand{\pl}{{{\cal P}_\infty}}
\newcommand{\thhalf}{{\scriptstyle{3/2}}}
\newcommand{\hhalf}{{\scriptstyle{1/2}}}
\newcommand{\half}{{\scriptstyle{\frac{1}{2}}}}

\newcommand{\refeq}[1]{\mbox{Eq.~(\ref{eq:#1})}}
\renewcommand{\k}{{\bf k}}

\newcommand{\basispl}{
   \put(-.5,-.5){\line(1,0){1}}
   \put(.5,-.5){\line(0,1){1}}
   \put(.5,.5){\line(-1,0){1}}
   \put(-.5,.5){\line(0,-1){1}}
                         }
\newcommand{\basisar}{
   \put(0,-.5){\vector(1,0){0}}
   \put(.5,0){\vector(0,1){0}}
   \put(0,.5){\vector(-1,0){0}}
   \put(-.5,0){\vector(0,-1){0}}
	              }
\newcommand{\plaq}{\setlength{\unitlength}{.5cm}\raisebox{-.2cm}{
   \begin{picture}(1.2,1.2)(-.6,-.6)
   \basispl\basisar
   \put(-.5,-.5){\circle*{.2}}
   \put(-.55,-.55){\makebox(0,0)[tr]{\footnotesize $x$}}
   \put(-.55,0){\makebox(0,0)[r]{\footnotesize $\nu$}}
   \put(0,-.55){\makebox(0,0)[t]{\footnotesize $\mu$}}
   \end{picture}}}

\newcommand{\twoplaq}{\setlength{\unitlength}{1cm}\raisebox{-.5cm}{
   \begin{picture}(1.2,1.2)(-.6,-.6)
   \basispl
   \put(-.5,-.5){\circle*{.1}}
   \put(-.5,.5){\circle*{.1}}
   \put(.5,-.5){\circle*{.1}}
   \put(.5,.5){\circle*{.1}}
   \put(0,-.5){\circle*{.1}}
   \put(0,.5){\circle*{.1}}
   \put(.5,0){\circle*{.1}}
   \put(-.5,0){\circle*{.1}}
   \put(-.25,-.5){\vector(1,0){0}}
   \put(.25,-.5){\vector(1,0){0}}
   \put(.5,-.25){\vector(0,1){0}}
   \put(.5,.25){\vector(0,1){0}}
   \put(-.25,.5){\vector(-1,0){0}}
   \put(.25,.5){\vector(-1,0){0}}
   \put(-.5,-.25){\vector(0,-1){0}}
   \put(-.5,.25){\vector(0,-1){0}}
   \put(-.55,-.55){\makebox(0,0)[tr]{\footnotesize $x$}}
   \put(-.55,0){\makebox(0,0)[r]{\footnotesize $\nu$}}
   \put(0,-.55){\makebox(0,0)[t]{\footnotesize $\mu$}}
   \end{picture}}}

\newcommand{\linkhmu}{\setlength{\unitlength}{.5cm}\raisebox{-.2cm}{
   \begin{picture}(1.2,1.2)(-.6,-.6)
   \put(.5,0){\line(-1,0){1}}
   \put(0,0){\vector(1,0){0.1}}
   \put(-.5,0){\circle*{.2}}
   \put(-.35,-.25){\makebox(0,0)[tr]{\footnotesize $x$}}
   \put(0.4,-.3){\makebox(0,0)[r]{\footnotesize $\mu$}}
   \end{picture}}}

\title{
Calorons with non-trivial holonomy on and off the lattice 
\vskip-3cm\hfill\small HU-EP-04/44, ITEP-LAT/2004-16\vskip2.6cm
}
\author{Falk Bruckmann\address{Instituut-Lorentz for Theoretical Physics, 
University of Leiden, P.O. Box 9506, NL-2300 RA Leiden, The Netherlands.},
E.-M. Ilgenfritz\address{Humboldt-Universit\"at zu Berlin, Institut 
f\"ur Physik, Newtonstr. 15, D-12489 Berlin, Germany},
B.V. Martemyanov\address{Institute for Theoretical and Experimental 
Physics, B. Cheremushkinskaya 25, 117259 Moscow, Russia},
M. M\"uller-Preussker${}^{\rm b}{}$, D\'aniel N\'ogr\'adi${}^{\rm a}{}$,\\
Dirk Peschka${}^{\rm b}{}$ and Pierre van Baal${}^{\rm a}{}$
\thanks{Presented by FB, EMI, DN and PvB at Lattice 2004.}
}
\begin{document}
\begin{abstract}
We discuss recent solutions for SU(2) calorons with non-trivial 
holonomy at higher charge, both through analytic means and using
cooling, as well as extensive lattice studies for SU(3).
\end{abstract}
\maketitle

\section{Introduction}\label{sec:Introduction}

We first briefly summarize the essential features of instantons at finite 
temperature $T$, so-called calorons~\cite{HaSh}. A non-trivial Polyakov 
loop at spatial infinity (the so-called holonomy), which in some sense 
plays the role of a Higgs field, reveals the monopole constituent nature 
of these calorons~\cite{Nahm,PLB,LeLu}. Trivial holonomy, i.e. with values 
in the center of the gauge group, is typical for the deconfined phase. 
Non-trivial holonomy is therefore expected to play a role in the confined 
phase ($T<T_c$) where the trace of the Polyakov loop fluctuates around zero. 
The properties of instantons are therefore coupled to the order parameter for 
the deconfining phase transition. 

Finite action requires the Polyakov loop at spatial infinity to be constant 
which for SU($n$) can be parametrized by
\beq
\pl=g^\dagger\exp(2\pi i\diag(\mu_1,\mu_2,\ldots,\mu_n))g,
\eeq
where $\sum_{m=1}^n\mu_m=0$ and $g$ is chosen to bring $\pl$ to its diagonal 
form, with the $n$ eigenvalues being ordered as $\mu_1\leq\mu_2\leq\ldots\leq
\mu_n\leq\mu_{n+1}\equiv1+\mu_1$. Caloron solutions are such that the total 
magnetic charge vanishes. A single caloron with topological charge one contains
$n\!-\!1$ monopoles with a unit magnetic charge in the $j$-th U(1) subgroup, 
which are compensated by the $n$-th monopole of so-called type $(1,1,\ldots,
1)$, having a magnetic charge in each of these subgroups~\cite{KvBn}. At 
topological charge $Q$ there are $|Q|n$ constituents, $|Q|$ monopoles of each 
of the $n$ types. The monopole of type $m$ has a mass $8\pi^2\nu_m/b$ and its 
core has a size $b/(4\pi\nu_m)$, with $\nu_m\!\equiv\!\mu_{m\!+\!1}\!-\!\mu_m$.
The sum rule $\sum_{m=1}^n\!\nu_m\!=\!1$ guarantees the correct action $8\pi^2
|Q|$ for calorons with topological charge $Q$. We can use the classical scale 
invariance to set $b\!=\!1/kT$, the period in the imaginary time direction, 
to 1.  

Perturbative fluctuations give rise to an effective potential as a function
of the background Polyakov loop, whose minima occur where the latter takes 
values in the center of the gauge group~\cite{Weiss}. Recently the 
non-perturbative contribution due to the calorons at non-trivial holonomy
was calculated~\cite{Diak}, and when added to the perturbative contribution, 
the minima at trivial holonomy turn unstable for decreasing temperature, 
right around the expected value of $T_c$. This lends some support to 
monopole constituents being the relevant degrees of freedom which drive 
the transition from a phase in which the center symmetry is broken at high 
temperatures to one in which the center symmetry is restored at low 
temperatures. 

The charge one SU($n$) action density, $\cS(x)=-\half\Tr F_{\mu\nu}^2(x)$, 
has a particularly simple form~\cite{KvBn},
\beqa
&&\hskip-6mm\Tr F_{\mu\nu}^{\,2}(x)\!=\!\partial_\mu^2\partial_\nu^2
\log\left[\half\tr(\cA_n\cdots \cA_1)-\cos(2\pi t)\right],\nonumber\\
&&\hskip-6mm\cA_m\equiv\frac{1}{r_m}\Biggl(\!\!\!\bea{cc}r_m\!\!&|\vec
y_m\!\!-\!\vec y_{m+1}|\\0\!\!&r_{m+1}\eea\!\!\!\Biggr)\Biggl(\!\!\!
\bea{cc}c_m\!\!&s_m\\s_m\!\!&c_m\eea\!\!\!\Biggr),\label{eq:dens}
\eeqa
with $r_m\!=\!|\vec x\!-\!\vec y_m|$ the distance to the $m^{\rm th}$ 
constituent monopole at $\vec y_m$, $c_m\!\equiv\!\cosh(2\pi\nu_m r_m)$, 
$s_m\!\equiv\!\sinh(2\pi\nu_m r_m)$, $r_{n+1}\!\equiv\! r_1$ and 
$\vec y_{n+1}\!\equiv\!\vec y_1$.  The chiral fermion zero mode is 
localized~\cite{ZM2,ZMN} to the $m^{\rm th}$ constituent monopole, 
provided one uses $\mu_m\!<\!z\!<\!\mu_{m+1}$ for the boundary condition 
$\hat\Psi_z(t+1,\vec x)\!=\exp(-2\pi iz)\hat\Psi_z(t,\vec x)$. ``Cycling'' 
through the values of $z$ gives the distinct signature of ``jumping"  
zero modes through which one can identify well-dissociated calorons, as 
observed in lattice studies for SU(2)~\cite{IMMV} and SU(3)~\cite{GatS}.

\begin{figure}[!b]
\vspace{3.0cm}
\includegraphics{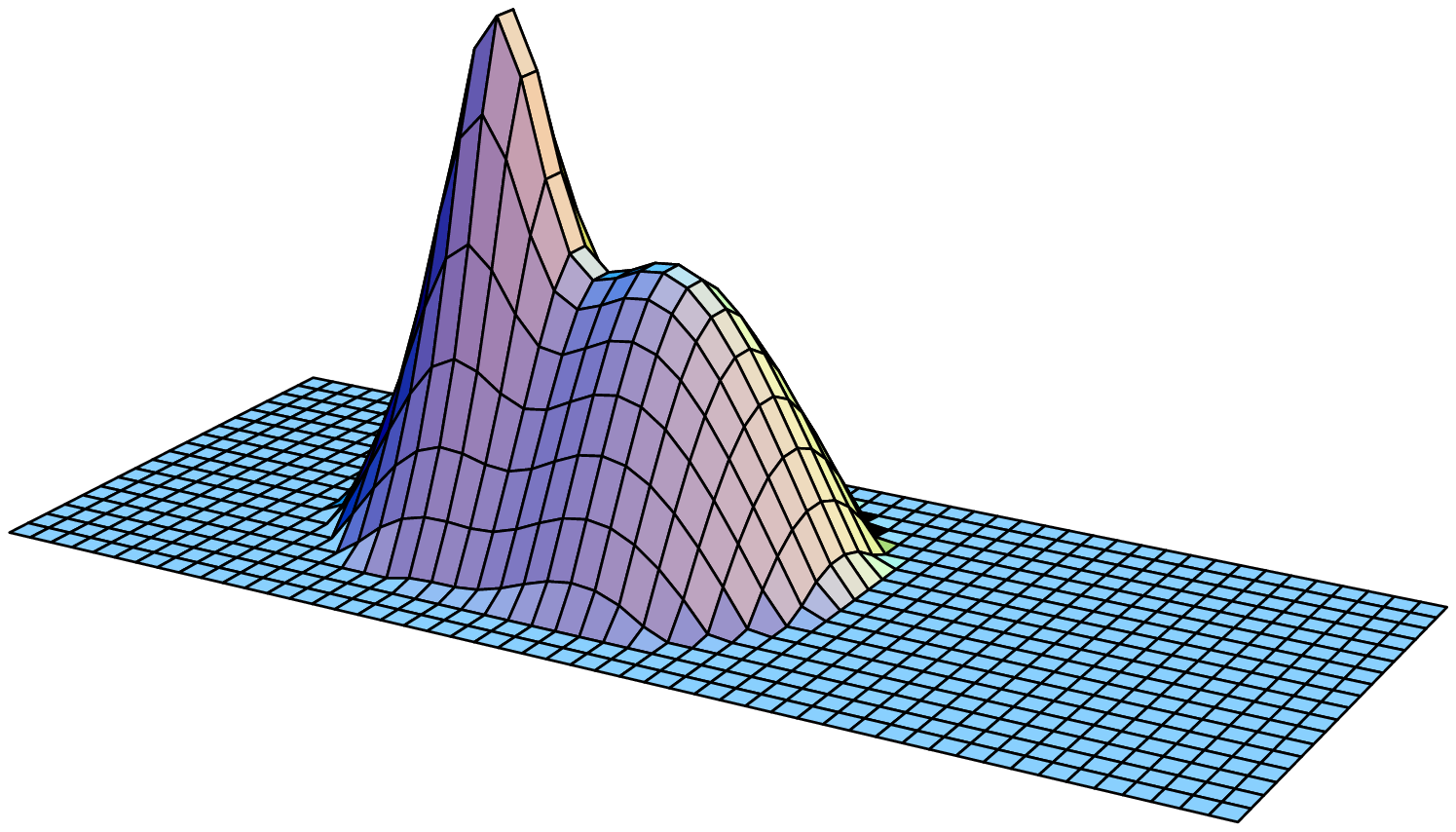}
\includegraphics{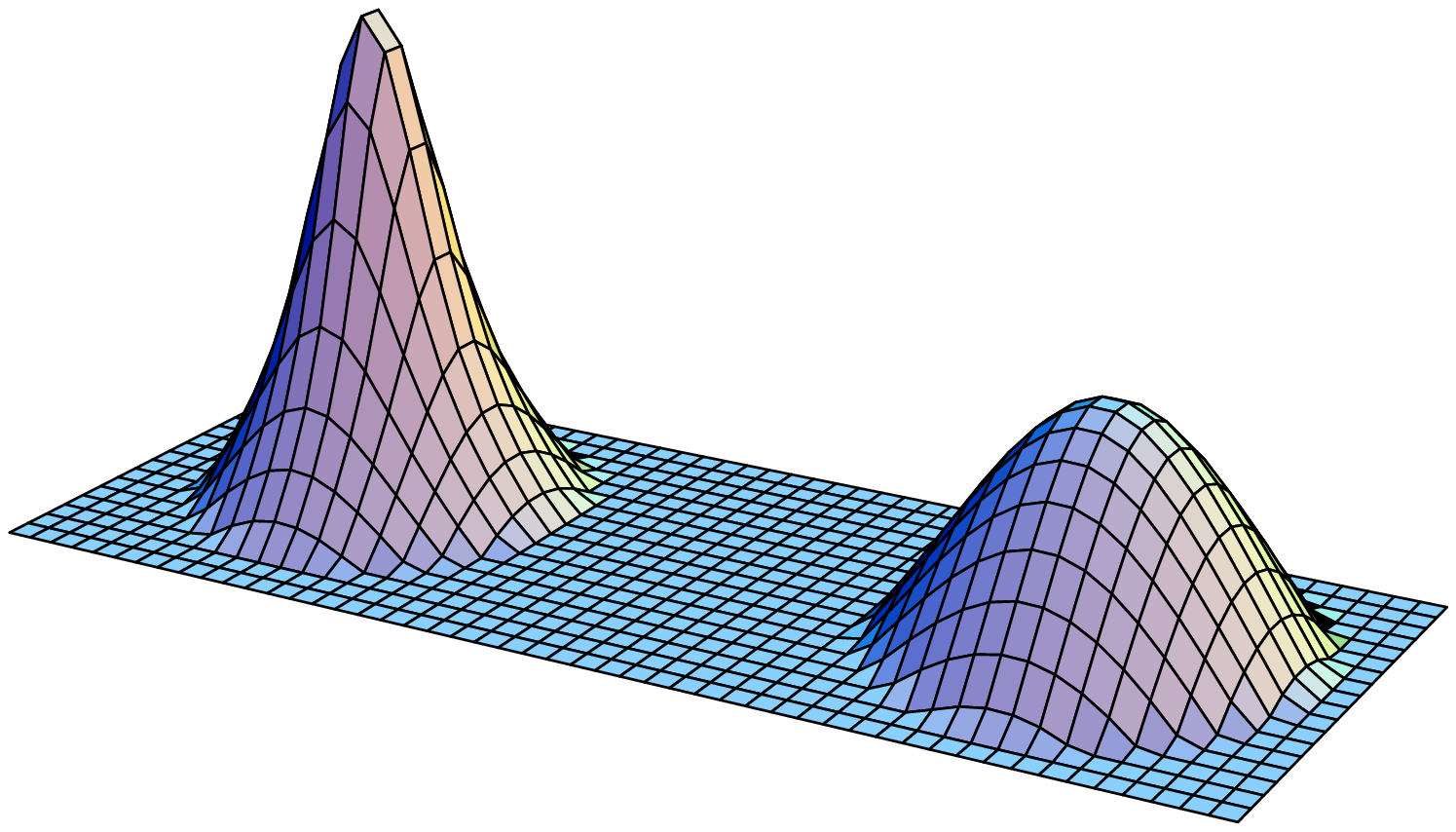}
\caption{Two charge one SU(2) caloron profiles at $t\!=\!0$ with 
$b\!=\!1$ and $\mu_2\!=\!-\mu_1\!=\!0.125$, for $\rho\!=\!1.6$ (bottom) 
and 0.8 (top) on equal lo\-ga\-rithmic scales (action density cutoff
at $1/(2e^2)$).}\label{fig:fig1}
\end{figure}
\begin{figure}[hbt]
\vspace{1.7cm}
\includegraphics{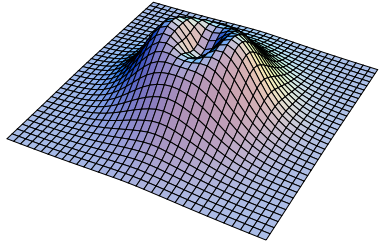}
\caption{The action density (on a linear scale) of a typical SU(2) caloron 
with topological charge 2 and $\mu_2\!=\!-\mu_1\!=\!0.25$ for which two 
constituents of equal magnetic charge are closer than their individual 
sizes (but not exactly on top). The other two constituents are far away, 
for which this becomes a static charge two monopole solution.
\vskip-9mm}
\label{fig:fig2}
\end{figure}

At any topological charge $Q$, well-separated constituents can be shown to act 
as point sources for the so-called far field, that is at large distance from 
any of the cores, where the gauge field is abelian~\cite{Us}. When constituents
of complementary charge ($n$ constituents of different type) come together, the
action density no longer deviates significantly from that of a standard 
instanton. Its scale parameter $\rho$ is related to the constituent separation 
$d$ through $\pi\rho^2/b=d$. A typical example for a charge one SU(2) caloron 
with far and nearby constituents is shown in Fig.~\ref{fig:fig1}. 
When $\rho\ll b$ no difference would be seen with the action density of the 
Harrington-Shepard solution~\cite{HaSh}, the gauge field is nevertheless 
vastly different, as follows from the fact that within the confines of the 
peak there are $n$ locations where two of the eigenvalues of the Polyakov loop 
coincide~\cite{PolM,PolP,PolI}. When, on the other hand, constituents of 
equal charge come together (which requires $|Q|>1$) an extended core structure 
appears~\cite{Us}. For two coinciding constituents this gives the typical 
doughnut structure also observed for monopoles~\cite{MonRing}, see 
Fig.~\ref{fig:fig2}.

\section{Analytic SU(2) results at higher charge}\label{sec:Analytic}

Also at higher topological charge the zero modes, denoted by $\hat\Psi_z^a(x)$,
where $a=1,\ldots,|Q|$, play an important role. We can write
\beq
\hat\Psi_z^a(x)^\dagger\hat\Psi_z^b(x)=-(2\pi)^{-2}
\partial_\mu^2\hat f^{ab}_x(z,z).\label{eq:zmdens}
\eeq
where $\hat f^{ab}(z,z')$ is a Green's function that appears in the 
construction to be discussed below. 

\subsection{The far field limit}

The trace, i.e. the sum over the zero-mode densities, has a remarkably simple 
form in the far field limit (denoted by $\ff$ and defined by neglecting terms 
that decay exponentially with the distance to any of the constituent cores)
\beq
\tr\hat f^\ff_x(z,z)=4\pi^2\cV_m(\vec x),\ \mbox{for}\ 
\mu_m<z<\mu_{m+1}.
\eeq
As is implicit in the notation, $\cV_m$ is static and independent of $z$ 
within its interval of definition. In addition $\cV_m$ has to be harmonic 
(up to singularities), because the zero modes decay exponentially as long 
as $z\neq\mu_j$ (for any $j$), and therefore do not survive in the far field 
limit. For $Q=1$ one simply has $\cV_m(\vec x)=1/(4\pi|\vec x-\vec y_m|)$, 
whereas for $Q=2$ we found~\cite{Us}
\beq
\cV_m(\vec x)\!=\!\frac{1}{2\pi|\vec x|}+\frac{\cD}{4\pi^2}\!\int_{r<\cD}
\hskip-5mm drd\varphi~\frac{\partial_r|\vec x-r\vec y(\varphi)|^{-1}}{
\sqrt{\cD^2-r^2}},
\eeq
where $\vec y(\varphi)\!=\!(\sqrt{1-\k^2}\cos\varphi,0,\sin\varphi)$, up to an
arbitrary coordinate shift and rotation. Here $\cD$ is a scale and $\k$ a shape 
parameter to characterize {\em arbitrary} SU(2) charge 2 solutions. In this 
representation it is clear that $\cV_m(\vec x)$ is harmonic everywhere except 
on a disk bounded by an ellipse with minor axes $2\cD\sqrt{1\!-\!\k^2}$ and 
major axes $2\cD$. Although not directly obvious, when $\k\to1$ the support 
of the singularity structure is on just two points, separated by a distance 
$2\cD$. Taking an arbitrary test function $f(\vec x)$ one proves that~\cite{Us} 
\beq
-\lim_{\k\to1}\int f(\vec x)\partial_i^2\cV_m(\vec x)d^3x=\sum_{z=\pm\cD}
f(0,0,z).
\eeq

\begin{figure}[b!]
\vspace{2.3cm}
\includegraphics{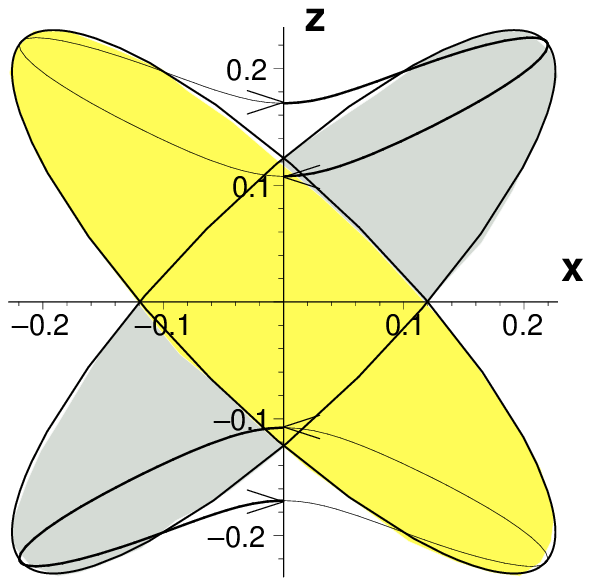}
\includegraphics{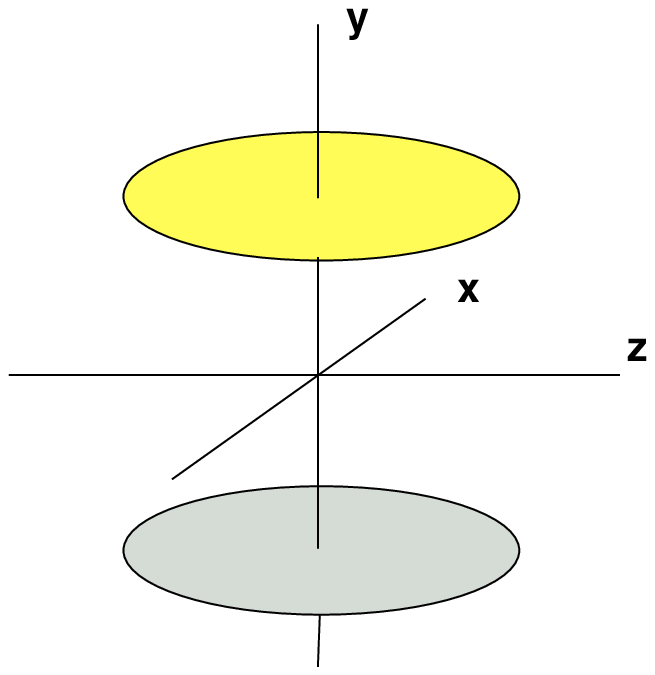}
\caption{Location of the disk singularities (shading according to magnetic 
charge) for a so-called ``rectangular" (left) and ``crossed" (right) 
configuration, as used for the $Q=2$ solutions shown in resp. 
Fig.~\protect\ref{fig:fig2} and Fig.~\protect\ref{fig:fig4}. Curves 
for the ``crossed" give a one parameter family of solutions interpolating 
between two axially symmetric solutions for which $\k=1$ independent of 
$\cD$ (see Ref.~\protect\cite{Us}).}\label{fig:fig3}
\end{figure}   
Monopoles of different charges have to adjust to each other to form an exact 
caloron solution, such that $\k$ and $\cD$ are in general not independent. So 
far we constructed two classes of solutions illustrated in Fig.~\ref{fig:fig3} 
for both of which a large value of $\cD$ implies that $\k$ approaches 1 
exponentially. Hence we find point-like constituents, a necessary requirement 
to describe the field configurations at larger distances in terms of these 
objects. When all constituents of other types are sent to infinity, we recover 
the exact multi-monopole solutions of a given type of magnetic charge, and our 
results therefore also provide explicit solutions for the monopole zero modes, 
which were not known before for the multi-monopole configurations.

Surprisingly, the charge distribution giving rise to the abelian field far 
from any of the constituent cores (even when extended due to overlap) can be 
calculated exactly from $\cV_m(\vec x)$. For SU(2) we can parametrize the 
holonomy by $\pl=\exp(2\pi i\vec\omega\cdot\vec\tau)$ ($\mu_2\!=\!-\mu_1\!=
\!|\vec\omega|$), where $\tau_a$ are the usual Pauli matrices. The field 
asymptotically becomes abelian, and is necessarily proportional to $\hat\omega
\cdot\vec\tau$. Hence $A^\ff_0(\vec x)=2\pi i\vec\omega\cdot\vec\tau-\half i
\hat\omega\cdot\vec\tau\Phi(\vec x)$, where the constant term shows again how 
$A_0$ plays the role of a Higgs field. For $Q=1$ and $2$ we have shown that 
$\Phi(\vec x)=2\pi(\cV_1(\vec x)-\cV_2(\vec x))$. Therefore, the singularity 
structure in the zero-mode density agrees {\em exactly} with the abelian 
charge distribution, as given by $\partial_i^2 \Phi(\vec x)$. Outside the 
cores there is no source for the abelian field, implying $A_0$ to be harmonic 
in the far field (as is also clear from the relation to $\cV_m$), and we find 
for the action density
\beq
\cS^\ff(\vec x)=-2\Tr(\partial_iA_0(\vec x))^2=-\partial_i^2\Phi^2(\vec x).
\eeq
Our expectation is that the result for $A_0^\ff(\vec x)$ will hold for 
arbitrary $Q$ and will generalize to SU($n$) with each $\cV_m(\vec x)$ 
simply associated to the U(1) generator with respect to which the magnetic 
charge of the monopole of type $m$ is defined.

\subsection{The construction -- in brief}

A charge $Q$ caloron with non-trivial holonomy is found with the
Atiyah-Drinfeld-Hitchin-Manin formalism~\cite{ADHM} by placing $|Q|$ instantons 
in the time interval $[0,b]$, which are repeated for each shift of $t$ over 
$b$, after a color rotation with $\pl$. The resulting infinite topological 
charge is dealt with by Fourier transformation, relating it to the Nahm 
transformation for calorons~\cite{Nahm}. The variable $z$ we introduced in 
formulating the generalized boundary conditions for the chiral fermions is 
precisely the dual of $t$ under this transformation. Singularities are 
introduced at $z=\mu_m$ through the powers of $\pl$, as is seen from writing 
$\pl=\sum_m e^{2\pi i\mu_m}P_m$ in terms of the $n$ projectors $P_m$. The 
Fourier transformation also produces $\hat A(z)$ as a U($|Q|$) gauge field 
on a circle, which satisfies the Nahm equation 
\beqa
&\!\!\!\!\!\ddz\hat A_j(z)+[\hat A_0(z),\hat A_j(z)]+\half\veps_{jk\ell}
[\hat A_k(z),\hat A_\ell(z)]\nonumber\\&\hskip2cm
=2\pi i\sum_m\delta(z-\mu_m)\rho_m^{\,j},\label{eq:nahm}
\eeqa
where $2\pi\zeta_a^\dagger P_m\zeta_b\!\equiv\!\Eins_2\hat S_m^{ab}\!-\!
\vec\tau \cdot\vec\rho_m^{\,ab}$, with $\zeta_b$ two-com\-po\-nent spinors in 
the $\bar n$ representation of SU($n$). Between the singularities $\hat A(z)$ 
is constant when $Q\!=\!1$, and for $Q\!=\!2$ it can be given~\cite{NahmM} 
in terms of the Jacobi elliptic functions ${\rm sn}_\k$, ${\rm cn}_\k$ and
${\rm dn}_\k$.

The Green's function $\hat f_x(z,z')$ (cmp. \refeq{zmdens}) is given by
$\hat g^\dagger(z)f_x(z,z')\hat g(z')$, defined through 
\beq
\left\{-\frac{d^2}{dz^2}+V(z;\vec x)\right\}\!f_x(z,z')=4\pi^2 
\delta(z\!-\!z')\Eins,\label{eq:green}
\eeq
with $V(z;\vec x)\equiv4\pi^2\vec R^2(z;\vec x)+2\pi \sum_m\delta(z-\mu_m)S_m$,
$R_j(z;\vec x)\equiv x_j-(2\pi i)^{-1}\hat g(z)\hat A_j(z)\hat g^\dagger(z)$
and $S_m\equiv\hat g(\mu_m)\hat S_m\hat g^\dagger(\mu_m)$. Note the latter
play the role of ``impurities". The ``dual" gauge function $\hat g(z)\equiv\exp
(2\pi i(\xi_0-x_0\Eins)z)$, defined in terms of the dual holonomy $\exp(2\pi 
i\xi_0)\equiv\Pexp(\int_0^1\hat A_0(z)dz)$, can be used to transform $\hat A_0
-2\pi ix_0\Eins$ to zero, in order to simplify as much as possible the 
Green's function equation. Although $\hat f_x(z,z')$ is periodic in $z$ 
and $z'$ with period $b$, $f_x(z,z')$ no longer is.

Given a solution for the Green's function, there are straightforward 
expressions for the gauge field~\cite{BrvB} only involving the Green's 
function evaluated at the ``impurity" locations and for the fermion 
zero modes~\cite{Us}. For the zero-mode density see \refeq{zmdens}. 
As an example we give the Green's function at $z'=z$, which formally 
can be expressed as (the $x$ dependence of $\cF_z$ is implicit) $f_x(z,z)
\!=\!-4\pi^2\left((\Eins\!-\!\cF_{z})^{-1}\right)_{12}$, where the 
$(1,2)$ component on the right-hand side of the first identity is with 
respect to the $2\!\times\!2$ block matrix structure, each of size
$|Q|\!\times\!|Q|$, and
\beq
\cF_{z}\!\equiv\!\hat g^\dagger(1)\Pexp\!\int_{z}^{z+1}\!\!dw
\pmatrix{0&\Eins\cr V(w;\vec x)&0\cr}.
\eeq
This has allowed us to find a compact formula for the action 
density~\cite{BrvB},
\beq
\Tr F_{\mu\nu}^2(x)\!=\!\partial_\mu^2\partial_\nu^2\log
\det\left(ie^{-\pi ix_0}(\Eins\!-\!\cF_{z})\right)\!,\label{eq:Sdens}
\eeq
which is actually independent of $z$ and reduces for Q=1 to \refeq{dens}.

The formal expression for $\cF_z$ can be simplified with $\cF_z\!=\!W_m(z)
W_m^{-1}(\mu_m)\cF_{\mu_m}W_m(\mu_m)W_m^{-1}(z)$, and decomposing $\cF_{\mu_m}$ 
into ``impurity" contributions at $z\!=\!\mu_m$ and the ``propagation" between 
them, resp. $T_m$ and $H_m\!\equiv\!W_m(\mu_{m+1})W^{-1}(\mu_m)$, with $T_m$,
$W_m(z)$ and $\cF_{\mu_m}$ given by
\beqa
&\hskip-4mm T_m\!\!\equiv\!\!\pmatrix{\!\Eins&\!\!\!\!0\cr\!2\pi S_m&\!\!\!\!
\Eins\cr}\!,\,W_m(z)\!\equiv\!\!\pmatrix{\ f_m^+(z)&f_m^-(z)\!\cr \!\ddz 
f^+_m(z)\!&\!\!\ddz f^-_m(z)\!\!\cr}\!,\!\!\!\!\!\!\nonumber\\&\hskip-15mm
\cF_{\mu_m}\!\equiv\!T_m\cdots H_1T_1\hat g^\dagger(1)H_n\cdots 
T_{m+1}H_m,\!\!\label{eq:build}
\eeqa
The matrices $f_m^\pm(z)$, defined for $z\!\in\!(\mu_m,\mu_{m+1})$, satisfy 
$f_m^\pm(z)\to\exp\left(\pm 2\pi |\vec x|(z-\mu_m)\right)f_m(\mu_m)$ 
for $|\vec x|\to\infty$. Together these form the $2|Q|$ solutions of the {\em 
homogeneous} Green's function equation. To resolve the full structure of the 
cores for $Q=2$ we had to find the exact solutions for $f_m^\pm(z)$ on each 
interval. For this task we could make convenient use of an existing analytic 
result for charge 2 monopoles~\cite{Hari}, adapting it to the case of calorons. 
Essential is that once $f_m^\pm(z)$ are known, everything else can be easily 
determined in terms of these, cmp. \refeq{build}. A sample of the 
results~\cite{Us} can be found in 
\begin{figure}[htb]
\vspace{3.6cm} 
\includegraphics{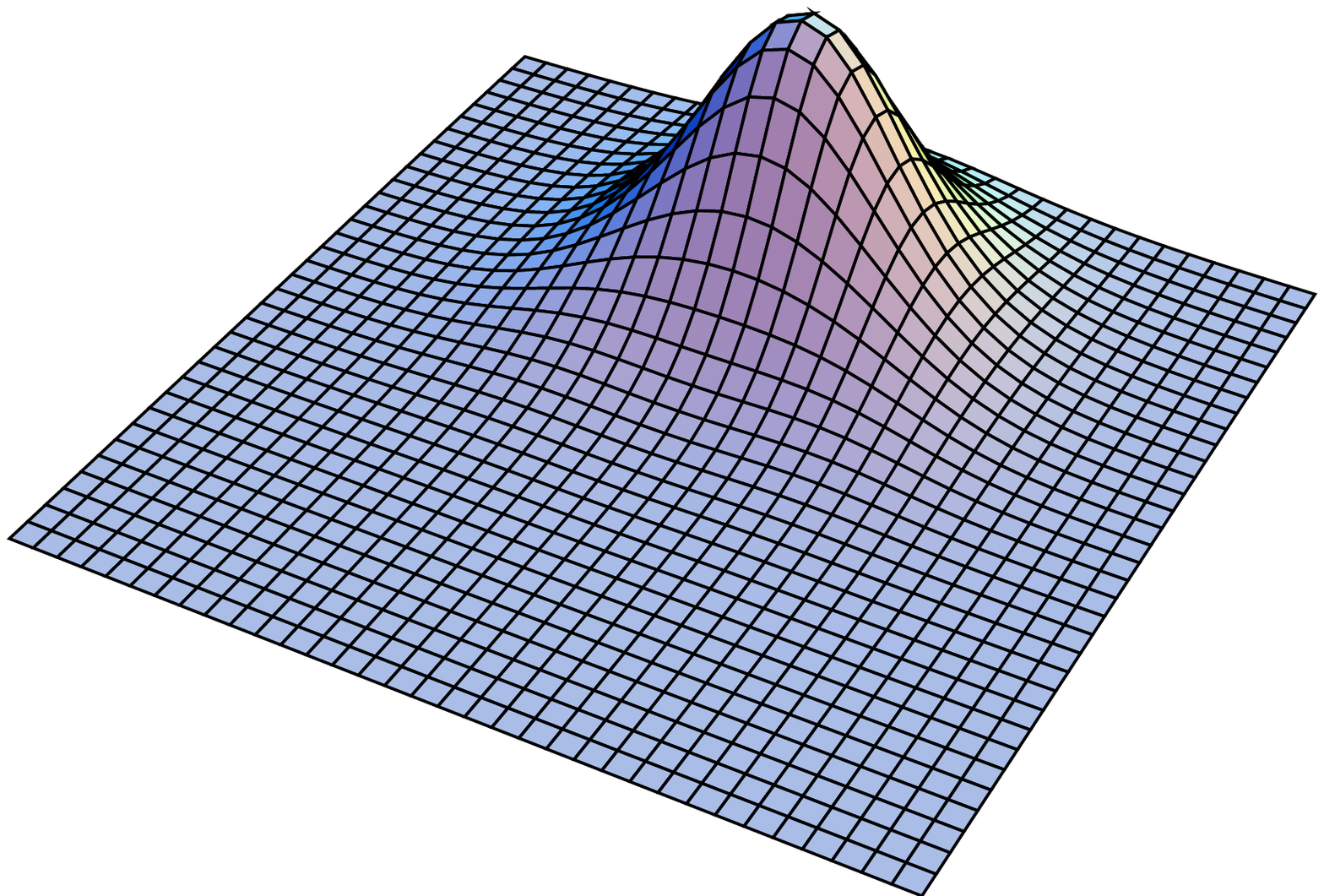}
\includegraphics{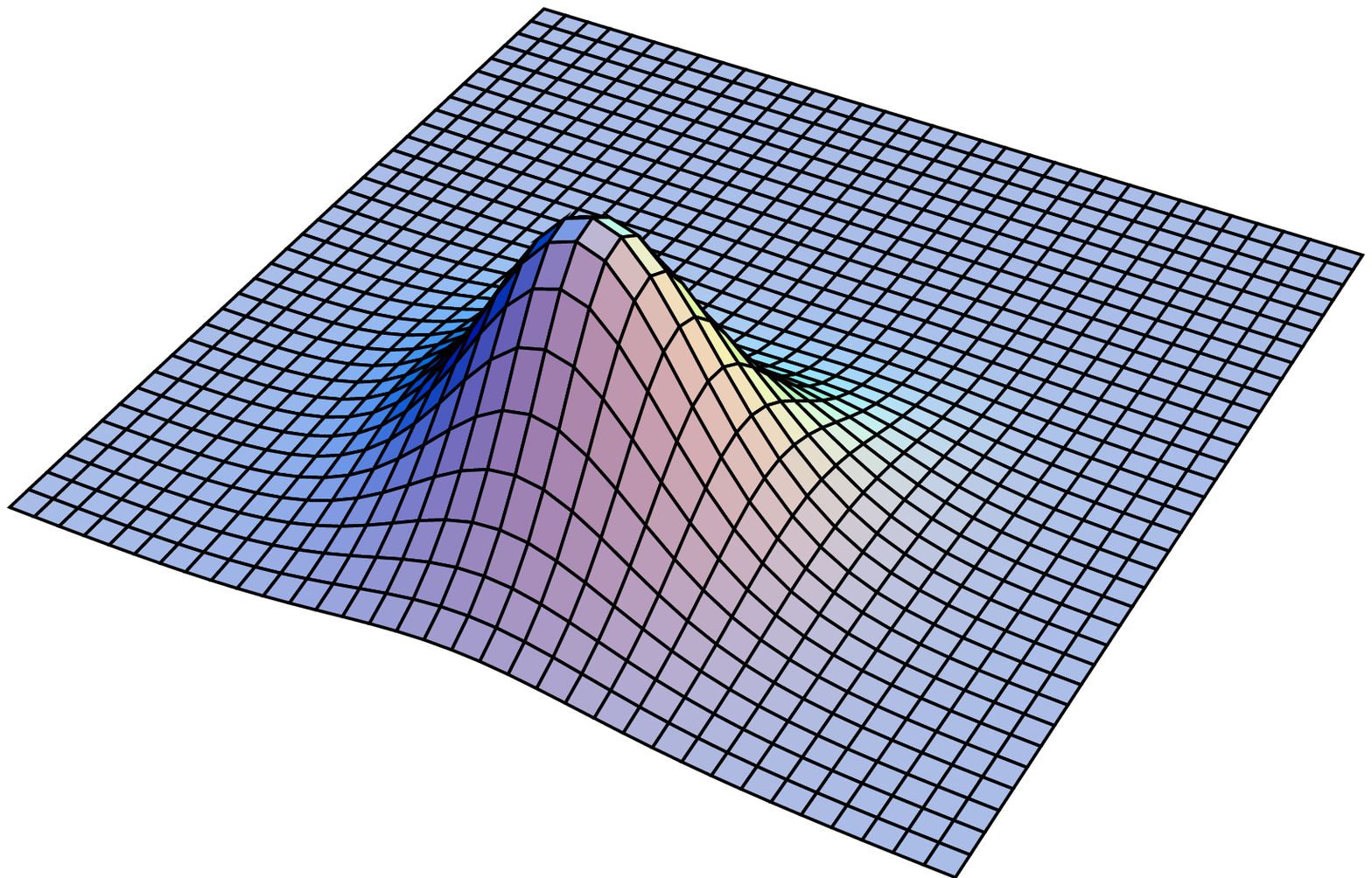}
\includegraphics{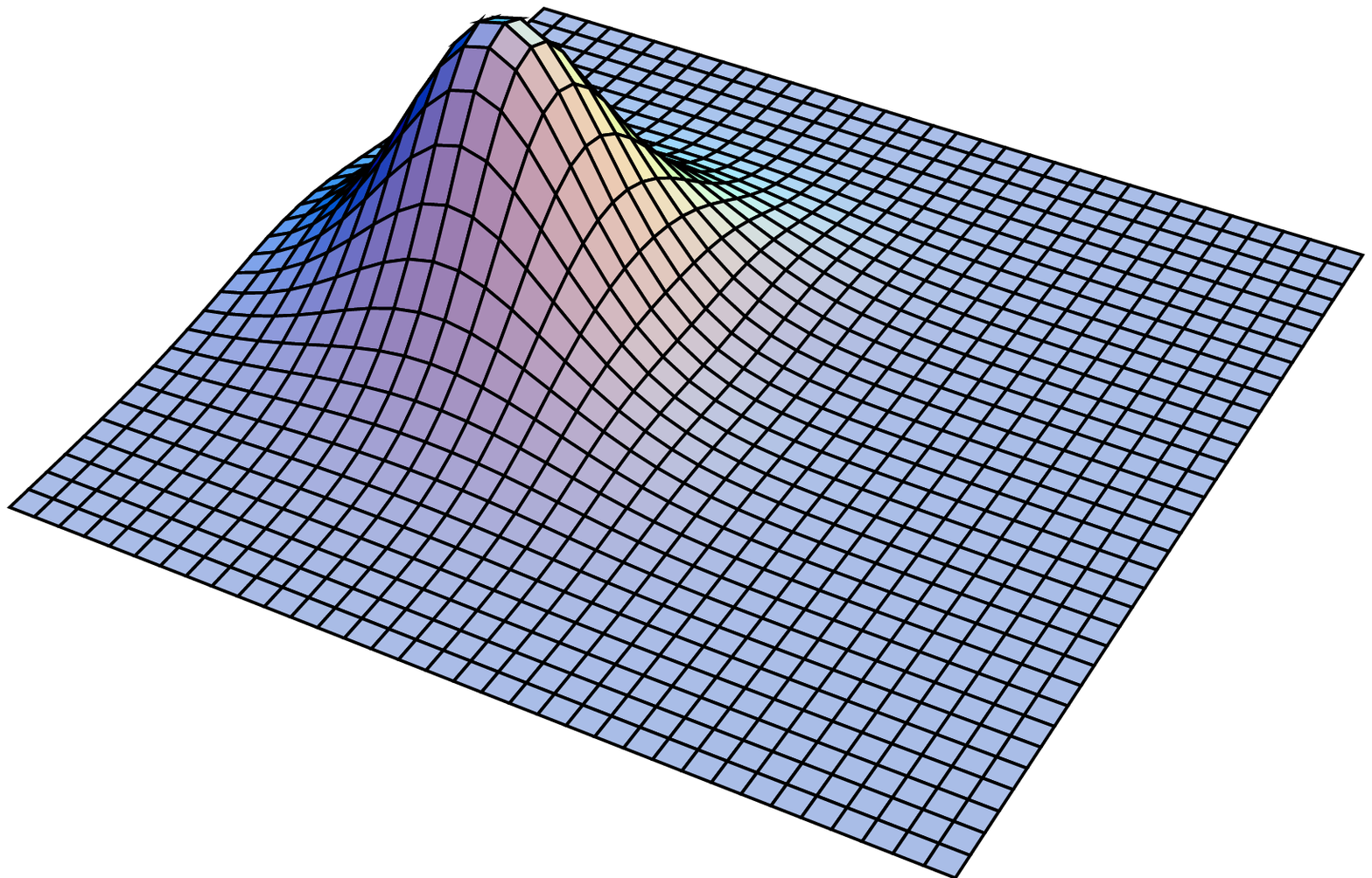}
\includegraphics{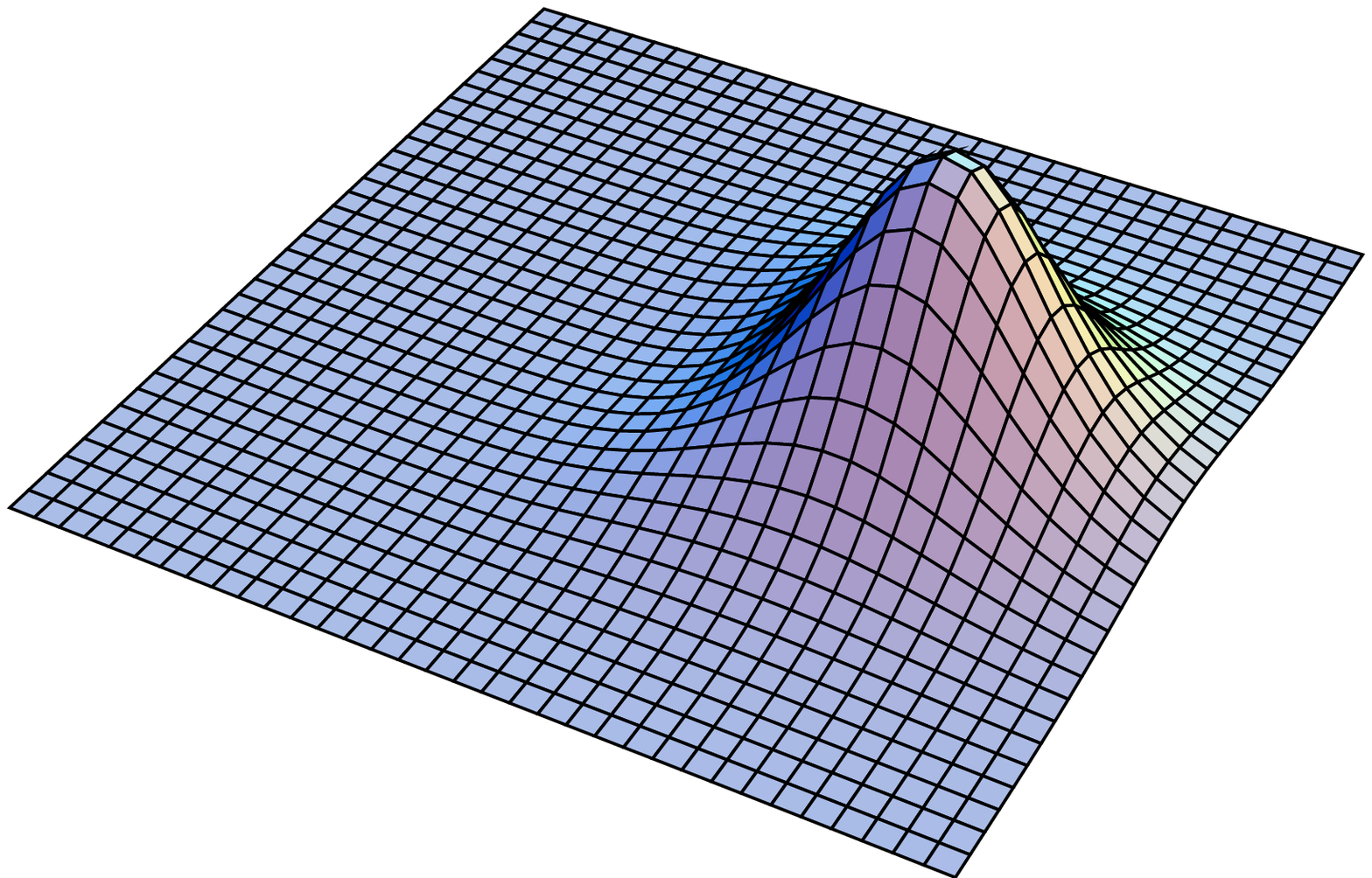}
\includegraphics{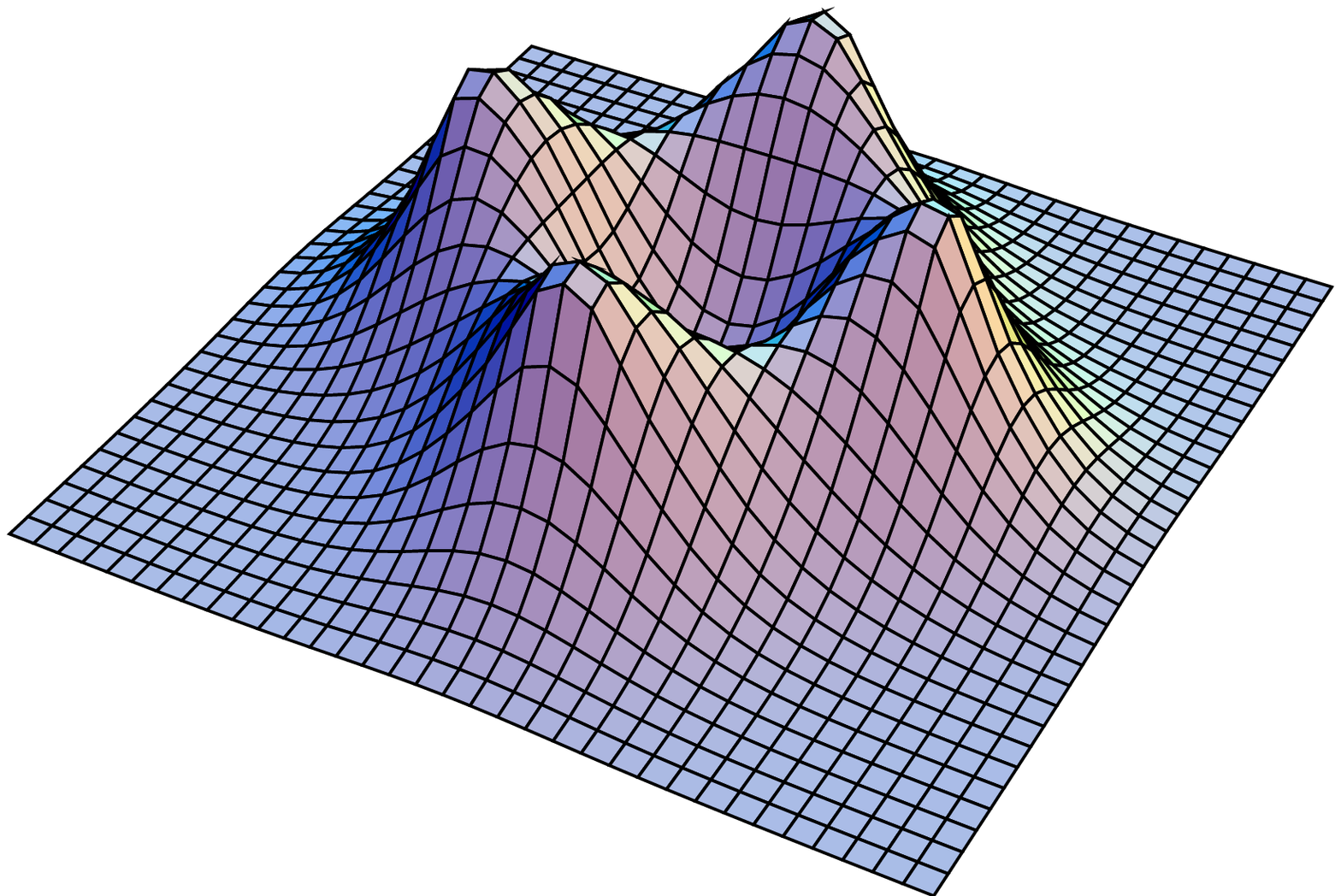}
\caption{In the middle is shown the action density in the plane of the
constituents at $t=0$ for an SU(2) charge 2 caloron with $\Tr\,\pl=0$ in the 
``crossed" configuration of Fig.~\ref{fig:fig3}. On a scale enhanced by a 
factor $10\pi^2$ are shown the densities for the two zero modes, using either 
periodic (left) or anti-periodic (right) boundary conditions in the time 
direction.
\vskip-9mm}
\label{fig:fig4} 
\end{figure}
Fig.~\ref{fig:fig2} and Fig.~\ref{fig:fig4}.

\section{Cooling studies for SU(2)}\label{sec:Cool}

Dissociated calorons in a dynamical setting were first observed for SU(2), 
using cooling to analyze Monte Carlo generated lattice configurations just 
below $T_c$~\cite{IMMV}. Yet, applying the same methods for a symmetric box, 
corresponding to low temperatures, no dissociation was seen~\cite{PolI},
in contrast to what was suggested by a fermionic zero-mode study~\cite{GatP}. 
Motivated by this we recently performed lattice cooling studies for SU(2), both
at finite and low temperatures, to analyze the constituent nature of self-dual 
configurations in more detail~\cite{Cool}. At finite temperature this can be 
compared with the analytic results, but such results are not available for 
the symmetric torus. We used so-called $\veps$-cooling~\cite{OvIm}, i.e. 
cooling with the improved action ($\linkhmu\equiv U_\mu(x)\in{\rm SU}(n)$) 
\beqa
S(\veps) & = & \sum_{x,\mu,\nu}\xi_\mu\xi_\nu\Biggl\{\frac{4-\veps}{3}
\,\mathrm{Re}~\Tr\left(\Eins-\plaq\right) \nonumber\\&+&\frac{\veps-1}{48}\,
\mathrm{Re}~\Tr\Biggl(\Eins-\twoplaq\Biggr)\Biggr\} \; .\label{eq:ecool}
\eeqa
The $\cO(a^2)$ lattice artifacts are proportional to $\veps$, and cause 
instantons to shrink for the Wilson action, which is recovered by putting 
$\veps=1$. This shrinking can be stabilized with $\veps=0$ or reversed with 
$\veps<0$, which was therefore called over-improvement~\cite{OvIm}. The 
$\xi_\mu$ are introduced to implement finite temperature (with $\xi_0=
\xi^\thhalf$ and $\xi_i=\xi^{-\hhalf}$) on a symmetric lattice, using 
an anisotropic coupling $\xi$. In this way the temperature can be lowered 
in arbitrarily small steps, after each of which we perform some cooling 
adjusting the configuration to a solution at this temperature. We called 
this process ``adiabatic" cooling. It allows us to start from a finite 
temperature configuration with well localized constituents, investigating 
how these evolve when lowering the temperature.

\subsection{From finite to ``zero" temperature}

\begin{figure}[!b]
\vskip4.7cm
\includegraphics{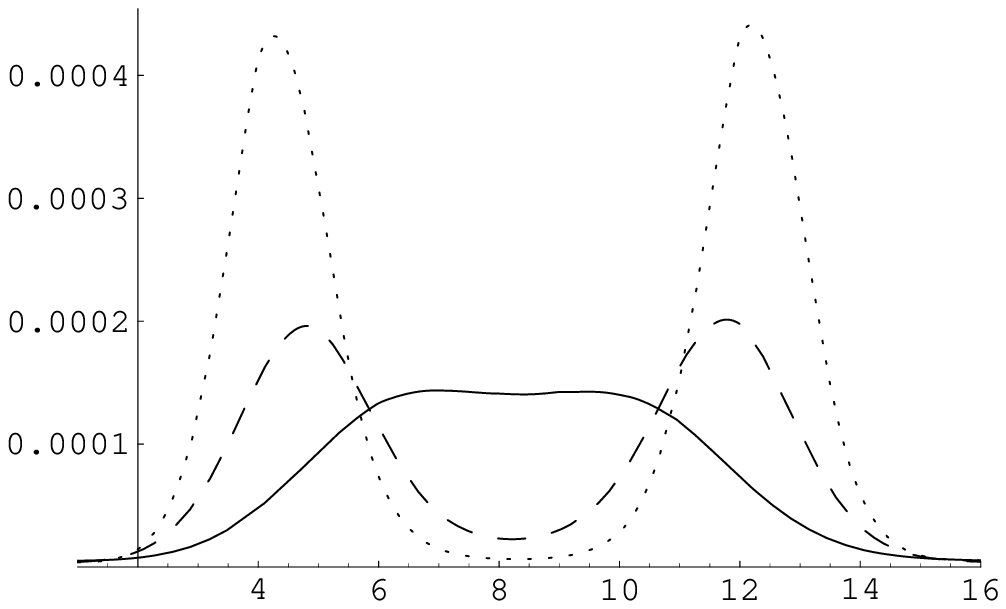}
\includegraphics{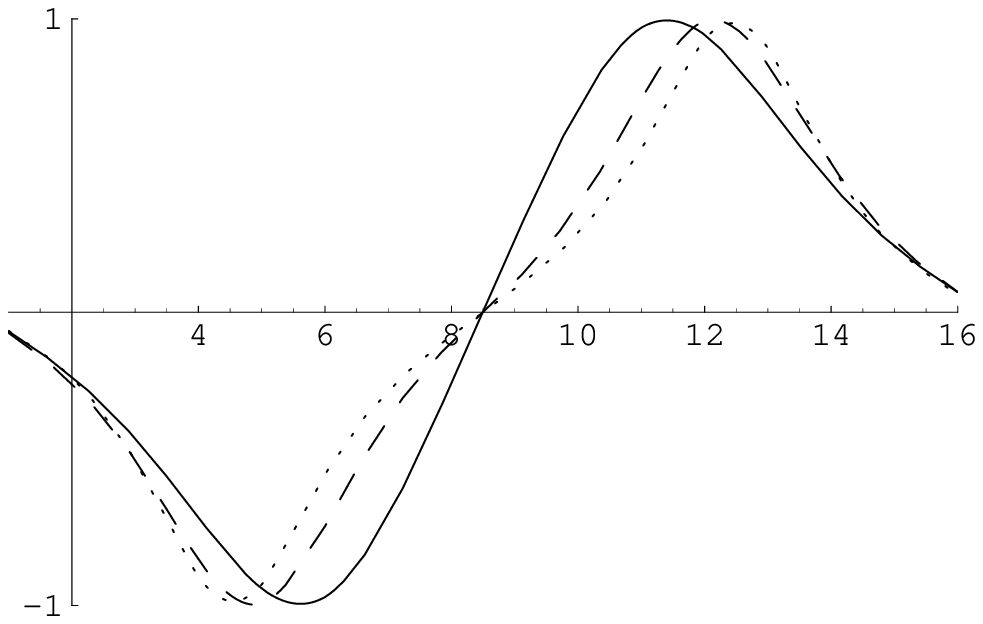}
\caption{Starting from a continuum caloron solution with well-separated
lumps, discretized on the anisotropic lattice and adjusted to the finite 
volume by 100 $\veps=-10$ cooling sweeps, we reduced $\xi$ from 4 to 1, 
through $\xi=2\sqrt 2$, 2 and $\sqrt 2$, applying between each of the 4 
steps 100 $\veps=-10$ cooling sweeps. Shown is the action density (top)
and the the Polyakov loop in the time direction (bottom) along a line 
through the constituent locations. The dotted, dashed and full curves 
are for $\xi=4$, 2 and 1.}\label{fig:fig5}
\end{figure}

Constituents called instanton quarks were conjectured to play a role 
long ago~\cite{InQu}. Based on our experience with calorons, it is expected 
that when lowering the temperature the constituents will grow in size. In a 
finite volume, as used on the lattice, the maximal constituent separation is 
limited. We conclude that when $b=L$, i.e. for the symmetric box (which 
will be denoted by ``zero" temperature), constituents necessarily overlap. 
In case of maximally non-trivial holonomy where all types of constituents 
are of equal mass, a periodic array of constituents is actually related
to twisted instantons of fractional topological charge~\cite{THoT,ToMa}. 
As soon as the constituents are not equidistant, they tend to 
merge and can no longer be identified from their action density profiles. 
In Fig.~\ref{fig:fig5} we present an example starting with two well-separated
constituents at finite temperature (using $\xi=4$ on a $16^4$ lattice),
lowering the temperature by reducing $\xi$ to 1.

\begin{figure}[!b] 
\vskip3.0cm
\includegraphics{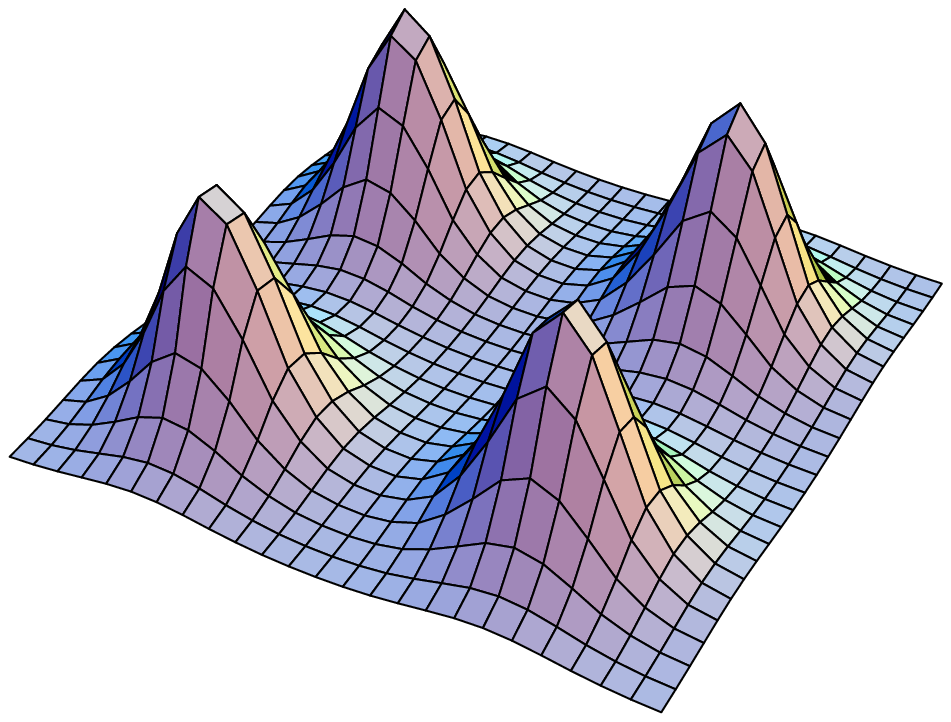}
\includegraphics{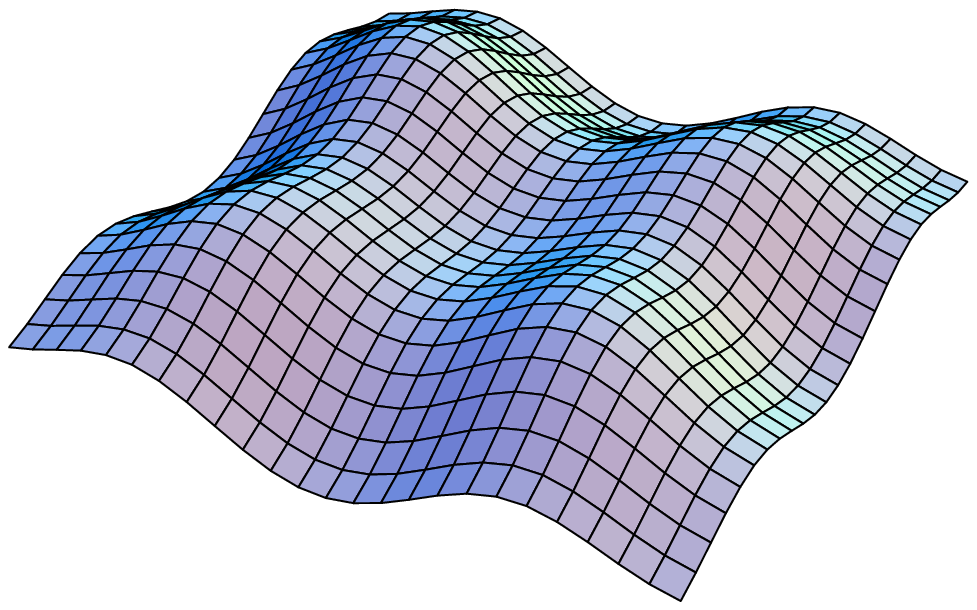}
\includegraphics{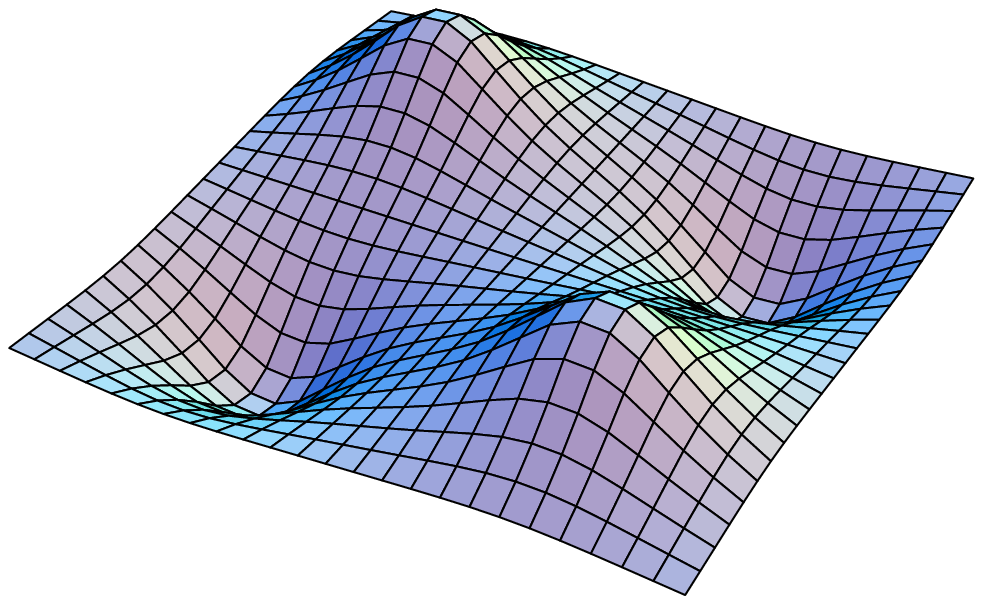}
\includegraphics{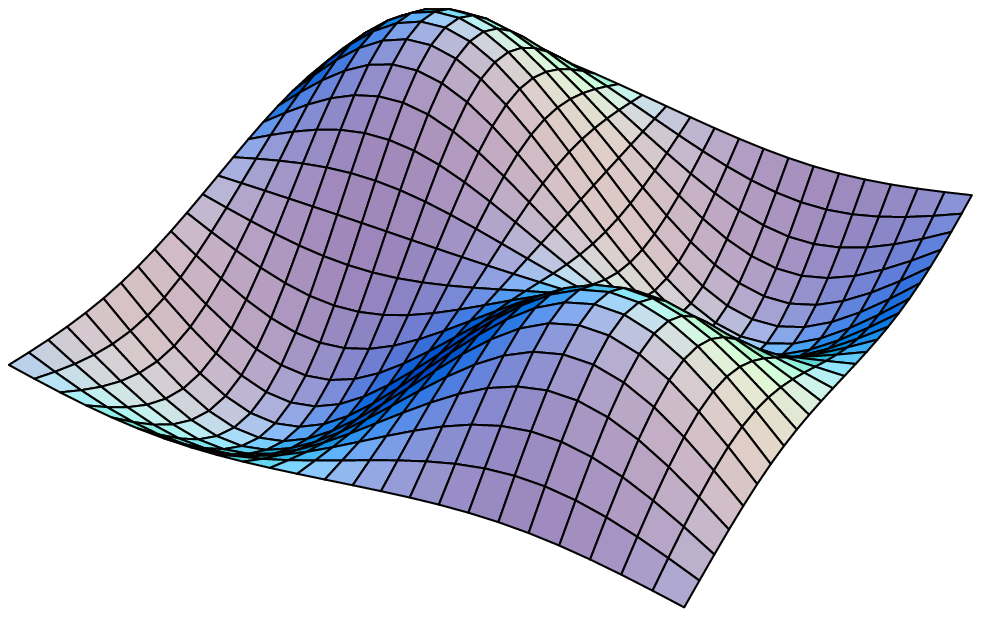}
\caption{Starting at finite temperature with a charge 2 caloron in the 
``crossed" configuration on a $16^4$ lattice with $\xi=4$ after 1000 cooling 
sweeps with $\veps=0$ (left two figures), changing $\xi$ through $2\sqrt2$, 
2, $\sqrt2$ to reach 1 (right two figures), at each of these applying 1000 
cooling sweeps with $\veps=0$. The top plots show the action density 
integrated over time, the bottom plots the Polyakov loop in the time 
direction, both in the $y$-$z$ plane at $x=8$ (where all constituents lie).}
\label{fig:fig6}
\vskip1.6cm
\includegraphics{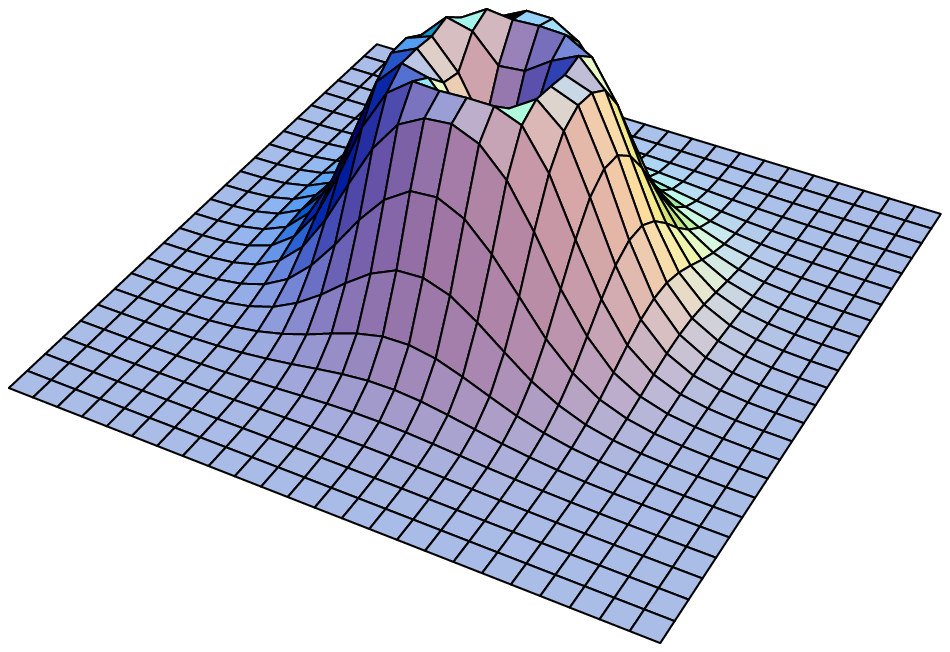}
\includegraphics{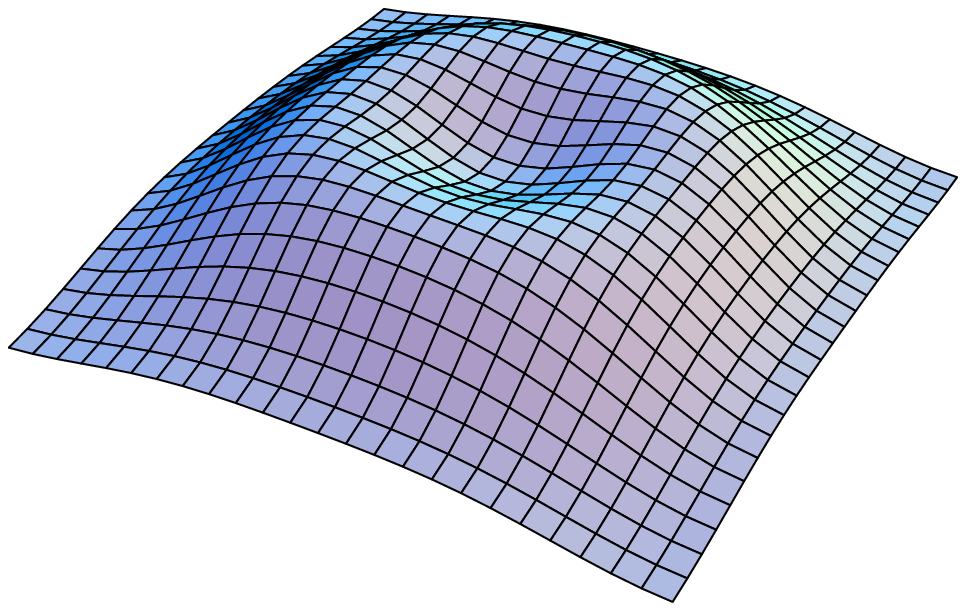}
\caption{As in Fig.~\ref{fig:fig6} but now starting from a ``rectangular" 
configuration first made at finite temperature ($\xi=4$, left) as described 
in the text. The action densities ($\xi=1$, right) integrated over time, 
is shown in the plane of the doughnut.}
\label{fig:fig7}
\end{figure}

There is one minor complication, that on a torus no charge one self-dual
solutions with finite size can exist, as can be proven with the help
of the Nahm transformation~\cite{Bra}. This means that charge 1 instantons
shrink (equivalent to the constituents getting closer) even when cooling 
with an action that has no lattice artifacts. This is why over-improvement
was used in Fig.~\ref{fig:fig5} to counteract this obstruction. In 
Figs.~\ref{fig:fig6} and \ref{fig:fig7} the result for the adiabatic 
cooling of a charge 2 caloron which is free from such obstructions
is shown. For Fig.~\ref{fig:fig6} we start at finite temperature with 
the so-called ``crossed" configuration (see Fig.~\ref{fig:fig3}, right). 

In Fig.~\ref{fig:fig7} the starting point at finite temperature (top) is 
the ``rectangular" configuration (see Fig.~\ref{fig:fig3}, left). It has 
$\Tr\,\pl=0$ and was taken from a Monte Carlo generated configuration 
at $\beta=2.2$, first cooling down to slightly above the 2 instanton 
action with $\veps=1$. After this, many thousands of $\veps=-2$ cooling 
sweeps were performed, followed by 500 with $\veps=0$. We only show the 
action density in the plane through one of the two doughnut structures,
the other doughnut is separated by half a period along the axis of symmetry. 
(The periodic zero modes are localized to one doughnut, and the anti-periodic 
zero modes to the other~\cite{Us,Cool}). To reach this configuration, we made 
use of the fact that under over-improved cooling constituents with opposite 
magnetic charge ``repel" whereas those with the same magnetic charge 
``attract", giving rise to constituents of equal magnetic charge to be exactly 
on top of each other. 

The adiabatic cooling is performed with $\veps=0$ cooling to prevent such 
``forces" due to the lattice artifacts while lowering the temperature. 
But when applying cooling with $\veps=-10$, the ``fat" doughnut (see 
Fig.~\ref{fig:fig7}, right) ultimately will reach the self-dual constant 
curvature solution that is allowed~\cite{THoC,PvBC} for a symmetric box 
with $Q=2$.

There is another way to avoid the $Q=1$ obstruction on a torus, namely by 
using twisted boundary conditions~\cite{THoT}, as was used for the study of 
calorons before~\cite{PolM}. Under much prolonged over-improved cooling with 
a twist in the $0$-$i$ planes we found somewhat to our surprise that it was 
possible to push the constituents further apart than half a period, at the 
same time inverting the magnetic charge of one constituent with respect to 
the other (this is in part due to the fact that the Polyakov loop is 
anti-periodic in the \mbox{$i$-th} direction due to the twist, for more 
details see Ref.~\cite{Cool}). This ultimately gave a perfect doughnut, 
i.e. two magnetic monopoles of {\em equal} charge on top of each other.

All these studies~\cite{Cool} have shown that at low temperatures 
constituents are not localized to 
better than half the size of the volume, both for the action density
and chiral zero modes. We conclude that well-localized lumps at zero 
temperature for low-charge self-dual backgrounds can only be found 
as instantons, even though it is clear that these are built from 
constituents of fractional topological charge, as seen through the 
behavior of the Polyakov loop. A good measure for the zero-mode 
localization~\cite{GatS} is the inverse participation ratio, $I\equiv 
V\sum_x\rho^2(x)$, where $\rho(x)$ is the zero-mode density
$\hat\Psi_z^\dagger(x)\hat\Psi_z(x)$. The bigger 
$I$ is, the more localized is the zero mode. This can be contrasted 
with a constant zero-mode density for which $I=1$. On average, at finite 
temperature~\cite{GatS} $I$ is indeed considerably larger than at zero 
temperature~\cite{GatP}, but in the latter case values of $I$ as big as 
20 or more are still seen to occur for cases where zero modes jump over 
distances as large as half the size of the volume when cycling through 
the boundary conditions. For our low-charge self-dual backgrounds the
zero modes associated with separated constituents never reached values 
of $I$ above~2.

The possibility that the zero modes are localized to instantons (formed 
from closely bound constituents) and jump between well-separated instantons, 
rather than well-separated isolated constituents, was discussed in 
Ref.~\cite{Tsu}. It was found that at finite temperature this is unlikely 
to occur, but it could not be ruled out for zero temperature. In a random 
medium of topological lumps the mechanism of localization of the zero modes 
could very well be similar to Anderson localization. In such a case one 
perhaps should expect a dependence on the fermion boundary conditions, even 
when constituents remain well hidden inside instantons. In the case that 
instantons form a dense ensemble, this is similar to the statement that it 
is impossible to determine which set of constituents forms an instanton. 
That some of the zero modes, if associated to fractionally charged lumps, 
are more localized than we observed~\cite{Cool} can have a dynamical origin. 

\subsection{Cooling histories}

That for temperatures just below $T_c$ well dissociated calorons are seen 
from the action density profiles has now been well established~\cite{IMMV}. 
Nevertheless, it is interesting that this can be deduced also from 
inspecting the cooling histories obtained with the Wilson action. 
\begin{figure}[!b]
\vskip1.1cm\hskip3.75cm$({\bf s})$\vskip1.2cm
\includegraphics{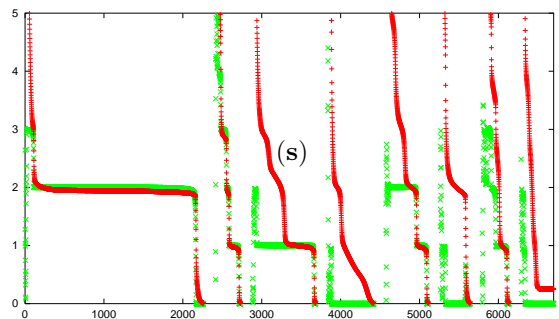}
\caption{Sample of 8 histories on a $16^3\times 4$ lattice
at $\beta=2.2$ ($T \approx 0.8 T_c$). Pluses give the Wilson action
and crosses the absolute value of the (order $a^2$ improved clover
averaged) topological charge.}\label{fig:fig8}
\end{figure}
There are three processes under which the action is lowered during cooling. 
The first and most important one is the removal of short distance fluctuations. 
Next is the annihilation of lumps with opposite topological charge, and 
last the falling of instantons through the lattice due to its shrinking. 
Annihilation can, however, also occur between well-localized lumps of 
opposite {\em fractional} topological charge. 

To detect these events one defines a plateau in the cooling history as the 
point of inflection for the action as a function of the cooling sweeps (i.e. 
when the decrease per step becomes minimal)~\cite{IMMV}. In the confined 
phase, where on average the charge fraction is one half, the change in action 
would typically be around one unit, and the topological charge would remain 
unchanged (an example is indicated by ``s" in Fig.~\ref{fig:fig8}). This can 
thus be easily distinguished from the annihilation of an instanton and 
anti-instanton, for which the action changes by two units. On the other 
\begin{figure}[htb]
\vskip-0.3cm
\hskip1.75cm\mbox{(i)}\hskip2cm\mbox{(ii)}\hskip1.9cm\mbox{(iii)}
\vskip4.0cm
\includegraphics{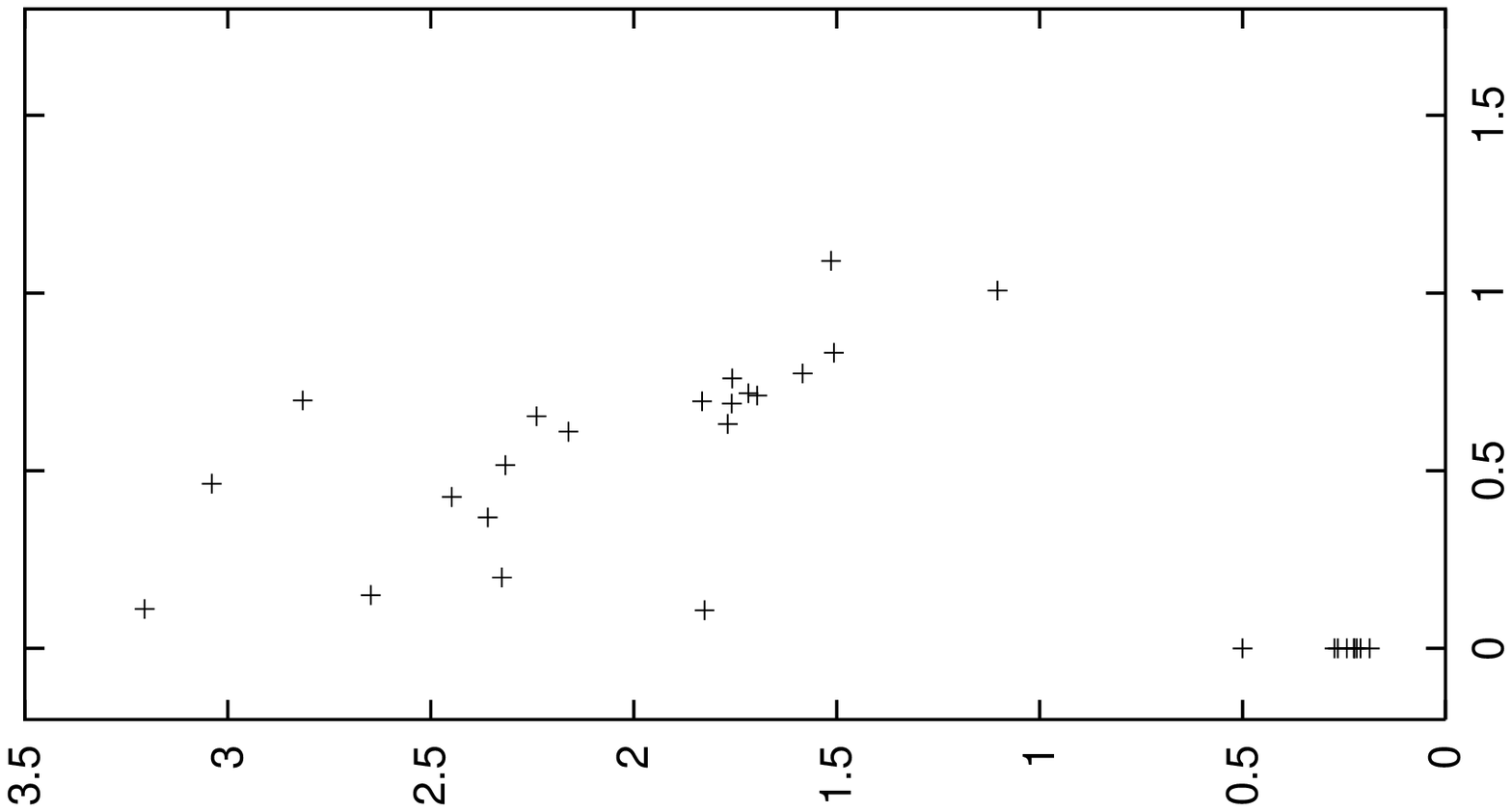}
\includegraphics{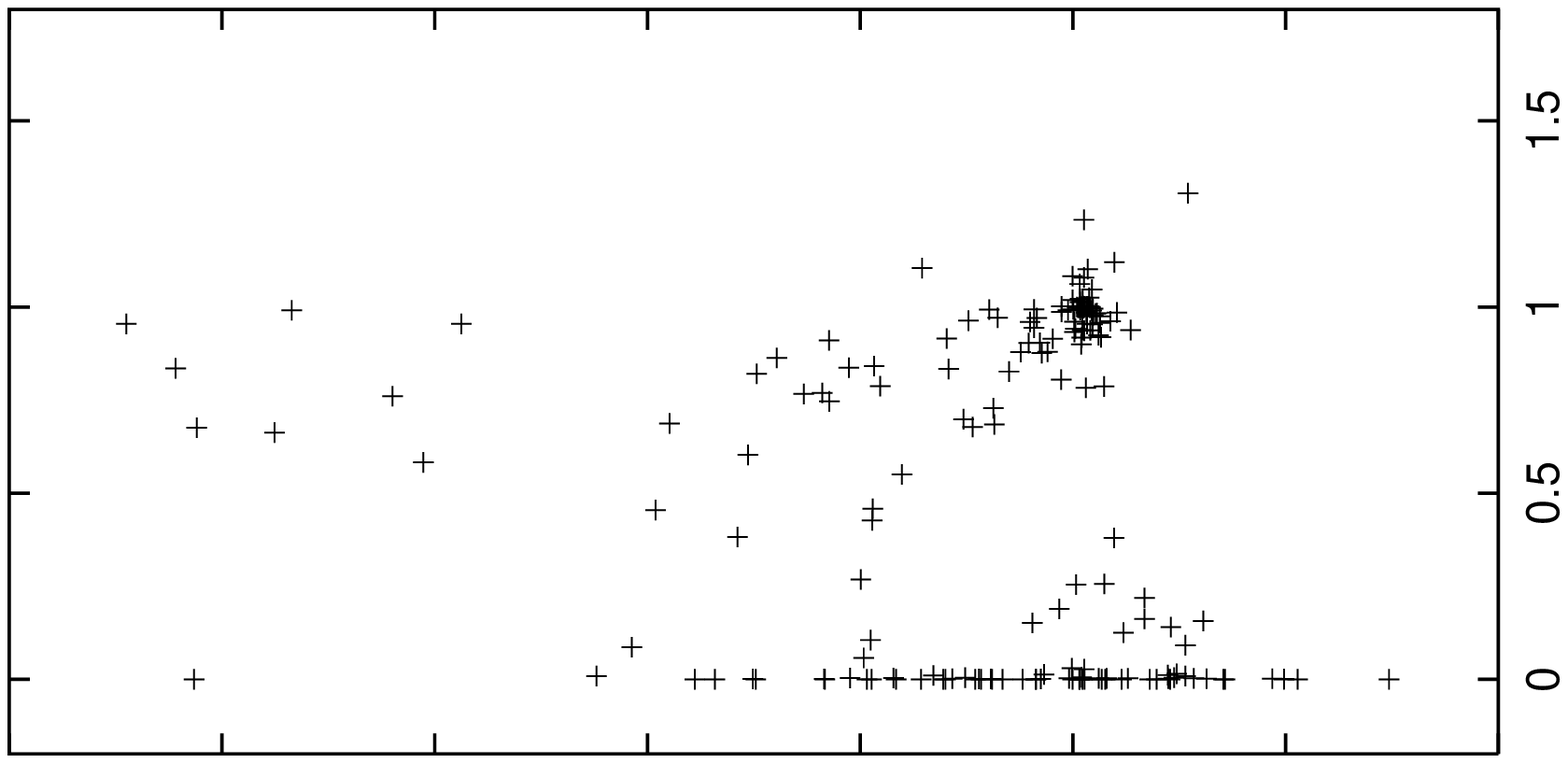}
\includegraphics{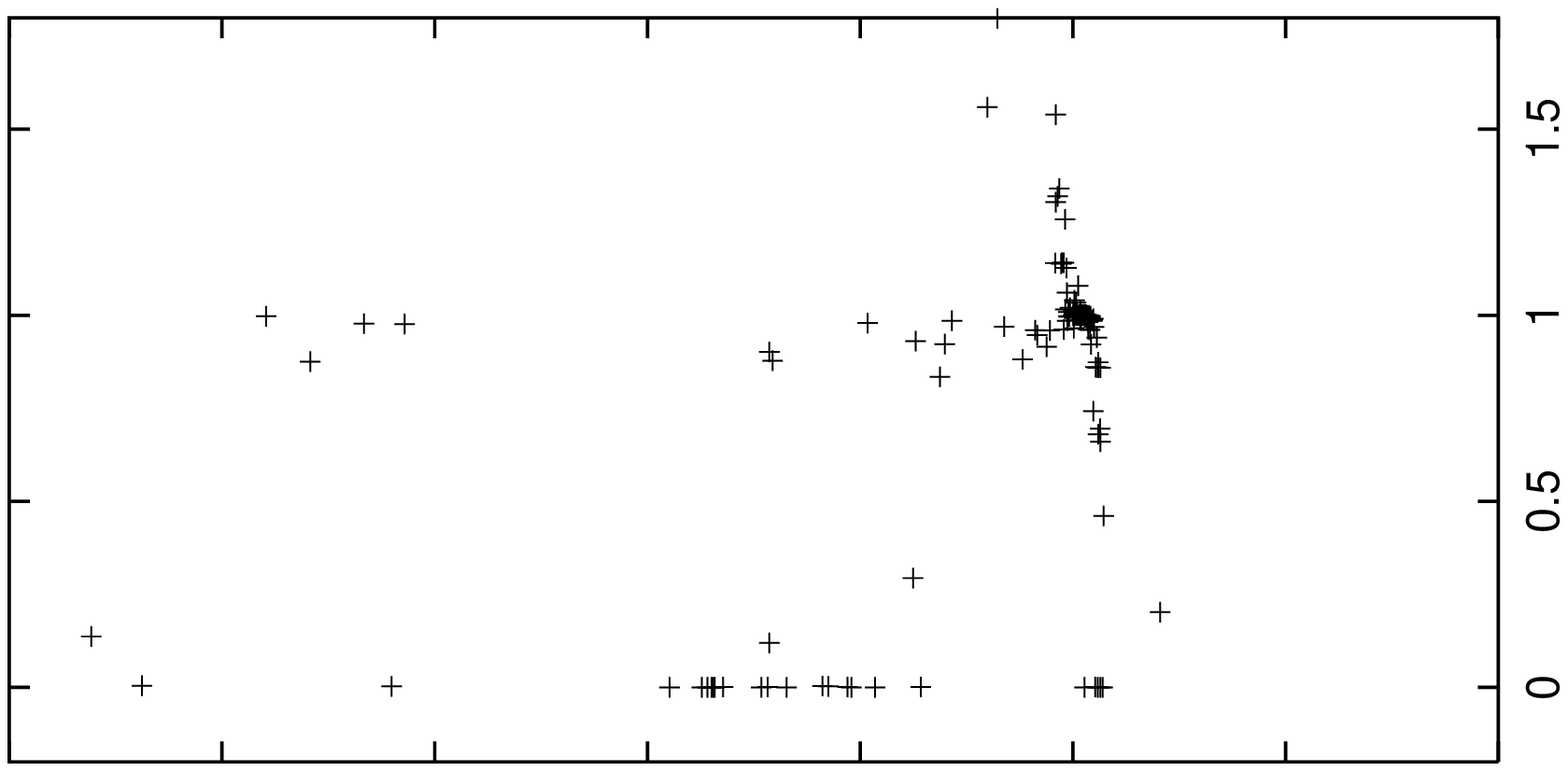}
\caption{Scatterplots of $|\Delta Q|$ (horizontally) versus $\Delta S$
(vertically) for a $6^3\times 4$ lattice at (i) $\beta=2.4$ ($T\approx 1.2
T_c$), (ii) $\beta=2.2$ ($T \approx 0.8 T_c$), and (iii) a $16^4$ lattice
at $\beta=2.3$ ($T\approx 0.25 T_c$).
\vskip-5mm}
\label{fig:fig9}
\end{figure}
hand, when an instanton falls through the lattice (i.e. two constituents 
with the same sign for their fractional topological charge, but of opposite 
magnetic and electric charge, come together and form a localized instanton, 
which then shrinks under cooling) both the topological charge and the action 
changes by one unit.

We collected the information concerning cooling histories in scatterplots, 
each based on 50 Monte Carlo generated configuration, see Fig.~\ref{fig:fig9}.
Only for $T$ just below $T_c$ the constituents are sufficiently well localized 
to give rise to clear constituent annihilations as seen from the 
clustering of $(\Delta S,|\Delta Q|)$ around $(1,0)$, measured from one 
plateau to the next. 

\begin{figure}[htb]
\vskip0.4cm\hskip0.6cm\hbox{Im$\,\lambda$}\vskip3.2cm
\hskip4.0cm\hbox{Re$\,\lambda$}\vskip-0.5cm
\includegraphics{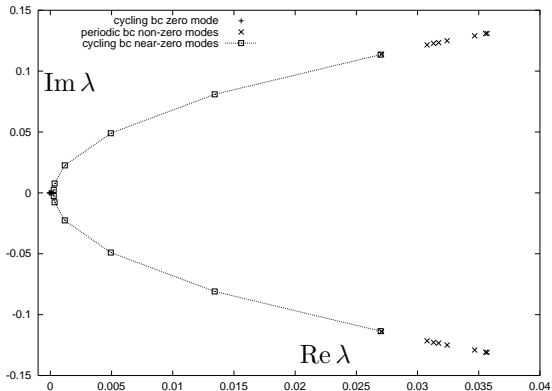}
\caption{Zero-modes for an anti-caloron on a $16^3\!\times\!4$ lattice, 
with in addition two constituents of opposite fractional topological charge.
Crosses give the low-lying eigenvalues $\lambda$ of the clover-improved 
Wilson-Dirac operator with periodic boundary conditions in time. The curves 
trace two near zero modes from their value with anti-periodic (left) 
boundary conditions to the periodic (right) case, in 5 equal steps 
in $z$ (squares).
\vskip-9mm}
\label{fig:fig10}
\end{figure}

The presence of constituents of opposite topological charge also gives rise 
to the interesting fact that the number of near zero modes (in a smooth, but 
non-selfdual background) can depend on the fermion boundary conditions as 
shown in Fig.~\ref{fig:fig10}. The anti-periodic zero modes localize to 
constituents for which the Polyakov loop is $-1$, whereas for periodic 
zero modes $P=+1$. The underlying configuration of Fig.~\ref{fig:fig10} has 
3 anti-selfdual lumps of which one has $P=+1$ and one self-dual lump with 
$P=-1$. Thus there are three anti-periodic near zero modes, two of which 
disappear as soon as $z\neq\half$.

\section{Marginally stable configurations}\label{sec:Marginal}

The last cooling run shown in Fig.~\ref{fig:fig8} ends in a configuration 
with $Q=0$ and $S\approx 1/4$ units, which is stable under further cooling. 
These so-called Dirac sheets had been shown to typically arise after 
constituent annihilation, leaving a constant magnetic field~\cite{DS1}.
It is perhaps surprising that such constant magnetic fields can be stable,
but this is a consequence of the finite volume in which a non-trivial
holonomy can affect the fluctuation spectrum. Since the action of the 
constant magnetic field does not depend on the holonomy, which also holds
on the lattice, this was called marginal stability~\cite{MaSt} and
explains the stability under further cooling, once it is reached. 

The finite temperature setting is essential, as otherwise one would need 
to introduce twisted boundary conditions to construct marginally stable 
configurations~\cite{MaSt}. For $SU(2)$ the constant magnetic field solutions 
are in fact abelian, with $F_{\mu\nu}=\pi i\tau_3n_{\mu\nu}/(L_\mu L_\nu)$
in a suitable gauge. Here $n_{\mu\nu}$ is antisymmetric, taking even integer 
values (in the absence of twisted boundary conditions). Adding a constant 
to the abelian vector potential can be partly, but not completely, absorbed 
by a translation. The part that cannot be absorbed will affect the fluctuation 
spectrum. With $L_i\!=\!L$, $n_{0\nu}\!=\!0$ and $m_i\!=\!\half\veps_{ijk}
n_{jk}$, the eigenvalues for the quadratic fluctuations can be shown 
to be~\cite{PvBC,MaSt} $\lambda_\pm\!\!=\!2\pi(2n+\!1\pm2)|\vec m|/L^2\!+
(2\pi e p+\!\vec m\cdot\vec C)^2\!/(|\vec m|L)^2\!+(2\pi q\!+\!C_0)^2/b^2$ 
and $\lambda_0=(2\pi k_\mu/L_\mu)^2$, with multiplicities resp. $2|e|$ and 
2, where $e$ is the greatest common divisor of the $m_i$. All quantum numbers 
($n,p,q,k_\mu$) are integer (but $n\geq 0$).

At finite temperature the spatial holonomies are typically trivial and the 
relevant Polyakov loop to consider for the dependence on the spectrum is 
$P_0=\half\Tr\exp(iC_0\tau_3/2)=\cos(C_0/2)$ or rather $P_0^2$. 
This is because the fluctuations involve adjoint fields and the spectrum 
is periodic under a shift of $C_0$ over $2\pi$, whereas $P_0$ is 
anti-periodic. To find the lowest eigenvalue we put $n\!=\!q\!=\!
p_i\!=\!C_i\!=\!0$ (we may restrict $|C_0|\leq\pi$) such that 
$\lambda_-=-2\pi|\vec m|/L^2+C_0^2/b^2$ is negative unless the Polyakov 
loop $P_0$ is sufficiently non-trivial. {\em All} eigenvalues are therefore 
positive when
\beq
|\vec m|<\frac{L\sqrt{\pi}}{b}\quad\mbox{and}\quad|P_0|<
\cos\left(\frac{|\vec m|b\sqrt{\pi}}{2L}\right).\label{eq:stab}
\eeq
These conditions can never be satisfied when $b=L$, nor in the deconfined 
phase where the Polyakov loop is too close to being trivial. We found in the 
cooling histories on $16^3\times4$ lattices stable configurations with one, 
two and three quarters of the instanton action (with one, two or three of the 
components of $\vec m$ being 2). For the minimal case, $|\vec m|=2$, the 
correlation between stability and the value of the Polyakov loop was studied 
in much detail~\cite{DS2}, showing perfect agreement with the analytic 
prediction in \refeq{stab}.

\section{Lattice calorons for SU(3)}

Caloron solutions with non-trivial holonomy for SU(3) are qualitatively 
different from those with trivial holonomy {\em and} from embedded SU(2) 
solutions, which appear only as limiting cases. We will illustrate this
with generic SU(3) solutions obtained by cooling on a finite lattice. 

\begin{figure}[!b]
\vskip2.5cm
\includegraphics{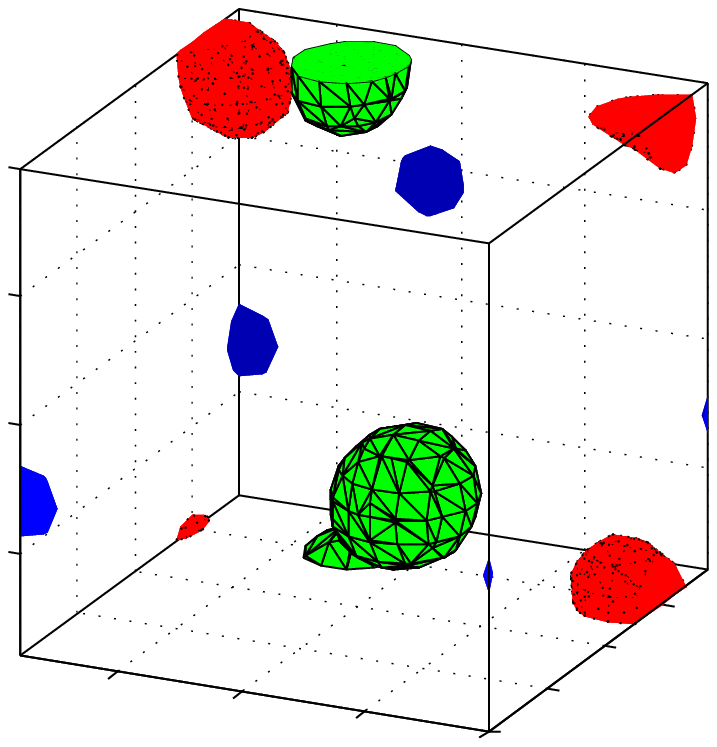}
\includegraphics{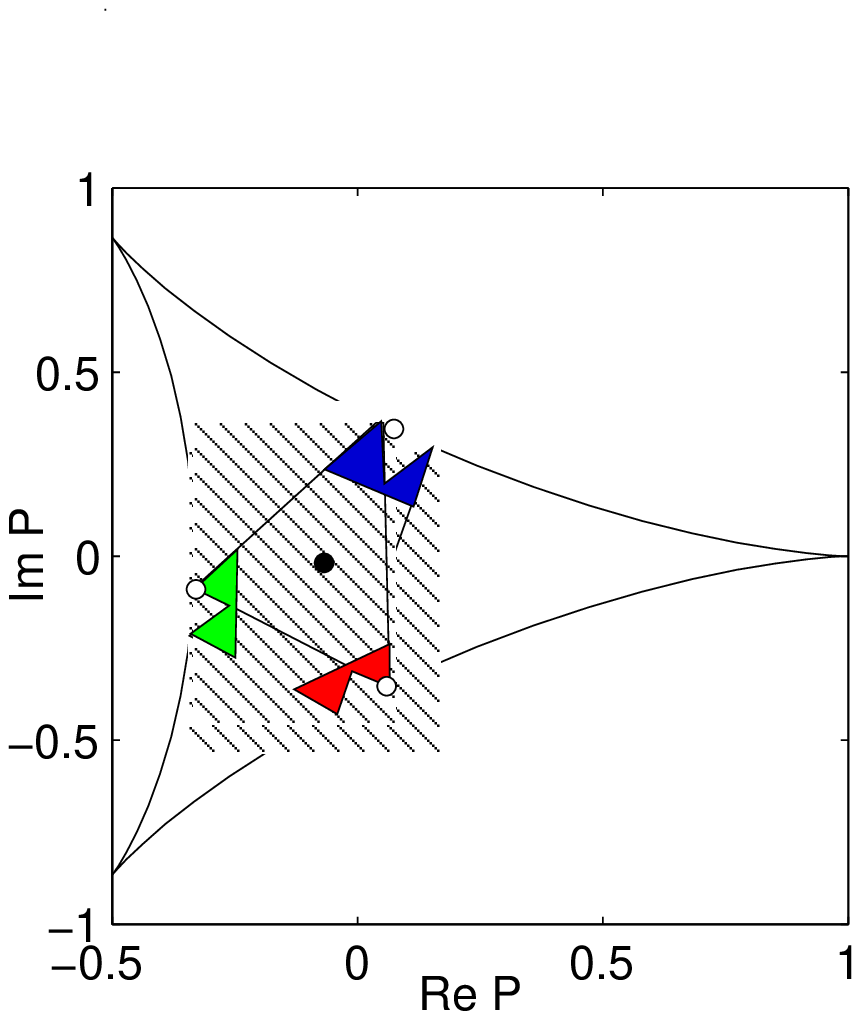}
\caption{Shown on the left by colored spheres are the six constituent 
monopoles for a $Q=2$ caloron. The behavior of the Polyakov loop is shown 
on the right, with the filled circle corresponding to the holonomy and the 
three open circles to the predicted values of $P$ at the constituent 
locations. The actual values of $P$ for points in their neighborhood 
are represented by the colored corners.}
\label{fig:polyakov_and_monopole}
\end{figure}

First we consider a $Q\!=\!2$ solution on a $20^3\!\times\!4$ lattice, 
see Fig.~\ref{fig:polyakov_and_monopole}.
Eigenvalues of the Polyakov loop all agree very well in the region where the 
action density is lowest (10 \% of the volume), and are used to define the 
holonomy $\pl$ in the case of a finite lattice. For calorons with one unit 
of topological charge it can be shown~\cite{PolP} that at the positions 
where two eigenvalues of the Polyakov loop coincide, these eigenvalues are 
given resp. by $\{e^{-\pi i\mu_3},e^{-\pi i\mu_3},e^{2\pi i\mu_3}\}$, $\{
e^{2\pi i\mu_1},e^{-\pi i\mu_1},e^{-\pi i\mu_1}\}$, and $\{-e^{-\pi i\mu_2},
e^{2\pi i\mu_2},-e^{-\pi i\mu_2}\}$. This is tested on the right in 
Fig.~\ref{fig:polyakov_and_monopole}. The typical range for $P$ (one-third
the trace of the Polyakov loop) is shown for two hypothetical $Q\!=\!1$ 
objects. It is spanned by the colored corners which correspond to the 
actual values inside the cores of the constituent monopoles. The colors 
distinguish the three different types of monopoles. For clarity we do not 
show the full scatterplot for $P$. The boundary for the range of $P$ values 
in the complex plane is indicated by the curved triangle. The closeness of 
two of the three eigenvalues is represented by the distance to this boundary. 
The six isosurfaces of that distance function are pictured in corresponding 
colors on the left in Fig.~\ref{fig:polyakov_and_monopole}, this is also 
where the action or topological charge density (not shown) is maximal. 

\begin{figure}[!b]
\vskip2.2cm
\includegraphics{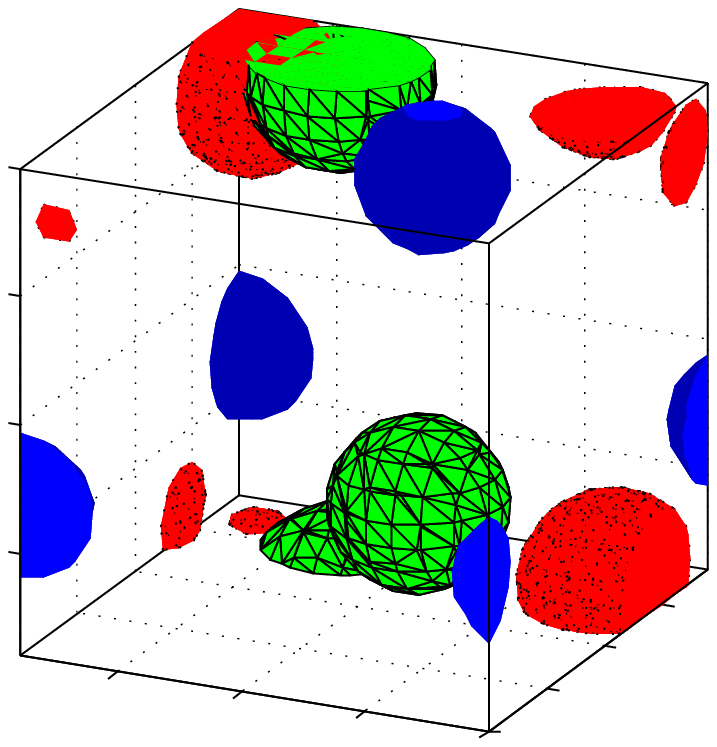}
\includegraphics{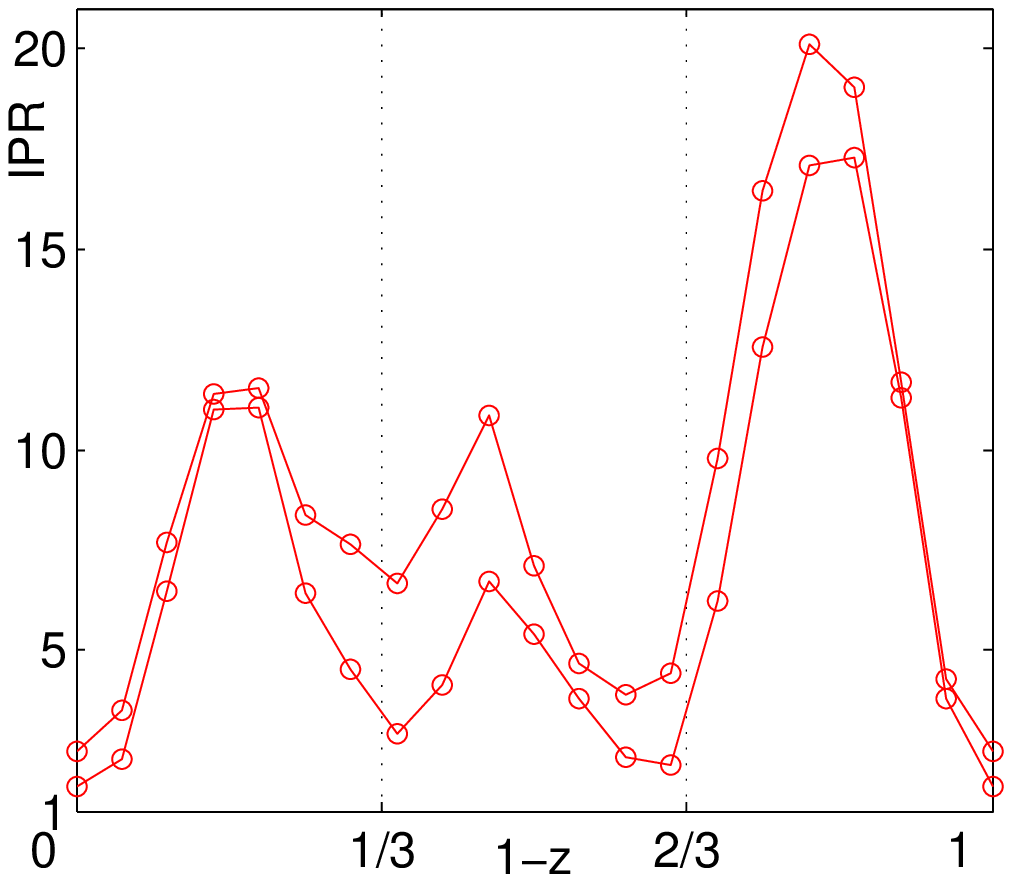}
\caption{The positions (red, blue, green) of the two zero modes (left) for 
the charge 2 configuration shown in Fig.~\ref{fig:polyakov_and_monopole} 
with the $z$ values corresponding to the maxima of $I$ (shown on the right).}
\label{fig:fermionic_and_ipr}
\end{figure}

The two zero modes for the charge 2 configuration of 
Fig.~\ref{fig:polyakov_and_monopole} are shown on the left of 
Fig.~\ref{fig:fermionic_and_ipr} as isosurfaces of its zero-mode density. 
Using colors we combine in one figure three values of $z$ that specify the 
fermion boundary condition in the time direction. These $z$ values correspond 
to the three maxima of the inverse participation ratio~\cite{GatS} $I$ (shown 
on the right), for which the zero modes are maximally localized. The monopole 
locations defined through the coinciding eigenvalues of the Polyakov loop 
agree (within one lattice spacing) with the maxima of the zero-mode density. 
With help of the $z$-dependent zero modes one can focus on one of the three 
types of monopoles alone, which brings out more clearly the constituent 
nature of calorons. 

Compared with the analytic caloron solutions with arbitrary holonomy for $S^1
\times\mathbb{R}^3$, lattice solutions have imperfections, some specific for 
SU(3). These are caused by the finite volume with periodic boundary conditions 
(leading to the charge one obstruction discussed in Sect.~\ref{sec:Cool}) and 
its finite aspect ratio $b/L$, as well as by lattice artifacts. Hence, several 
compromises are necessary to describe the moduli space of caloron solutions 
using {\em cooling on the lattice}. In this spirit we have performed a 
systematic investigation~\cite{in_preparation} of the structure of SU(3) 
calorons with $|Q|=1, 2, 3$ and 4 on $N_s^3\times N_t$ lattices with $N_s=12$ 
and 20 ($N_t \geq 4$). For $N_s=12$ the results are based on $\cO(10000)$ 
cooled configurations. The holonomy is non-trivial when applying cooling to 
configurations generated in the confined phase using $\beta=5.6$ 
(corresponding to $T\approx 0.95 T_c$ for $N_t=4$). Similar to what was seen 
for SU(2)~\cite{IMMV}, the distribution of the Polyakov loop broadens with 
cooling, and cuts in $P_{\infty}\!=\!\Tr\,\pl/3$ are sometimes helpful to 
highlight the constituent nature. 

For an automated scanning of the solutions we introduced two observables, the 
non-staticity~\cite{IMMV} $\delta_t\!\equiv\!\frac{N_t}{4}\sum_x\left|s(x)-\!
s(x\!+\!\hat{t})\right|/S$ and the violation of (anti)self-duality $\delta_F
\!\equiv\!\sum_x\left|s(x)\!-\!8\pi^2|q(x)|\right|/S$. The non-staticity is 
used to determine when separated lumps are seen within a solution. Points of 
bifurcation occur when $\delta_t^{*} =0.27$ and $\delta_t^{**}=0.06$ for which 
respectively two and three peaks become visible in the topological charge 
density. These values were estimated on the basis of the exact charge one 
caloron solutions with maximally non-trivial holonomy and suitably arranged 
constituents~\cite{in_preparation}. The $\cO(a^4)$-improved field 
strength~\cite{improv_fieldstrength} was used to construct the action and 
topological charge densities, to minimize the effect of lattice artifacts. 
For the fermions with adjustable boundary conditions we used the 
clover-improved Wilson-Dirac operator.\vskip-.1mm

Like for SU(2), instantons and calorons shrink under cooling with the Wilson 
action. In addition, charge one caloron solutions can only be obtained at the 
expense of violation of (anti)self-duality. For example, cooling on a $16^3\! 
\times\!4$ lattice with a stopping criterion based on a minimal violation of 
the lattice equations of motion~\cite{IMMV}, or a minimal value of $\delta_F$, 
one is unable to produce dissociated $|Q|=1$ calorons, although this can 
be corrected for using over-improved cooling, as we have seen for SU(2) 
in Sect.~\ref{sec:Cool}. However, for our high-statistics study we have 
restricted ourselves to Wilson-type cooling. Instead we modified the condition 
under which cooling is stopped and the approximate solution is stored. 

\begin{figure}[!b]
\vskip5cm
\includegraphics{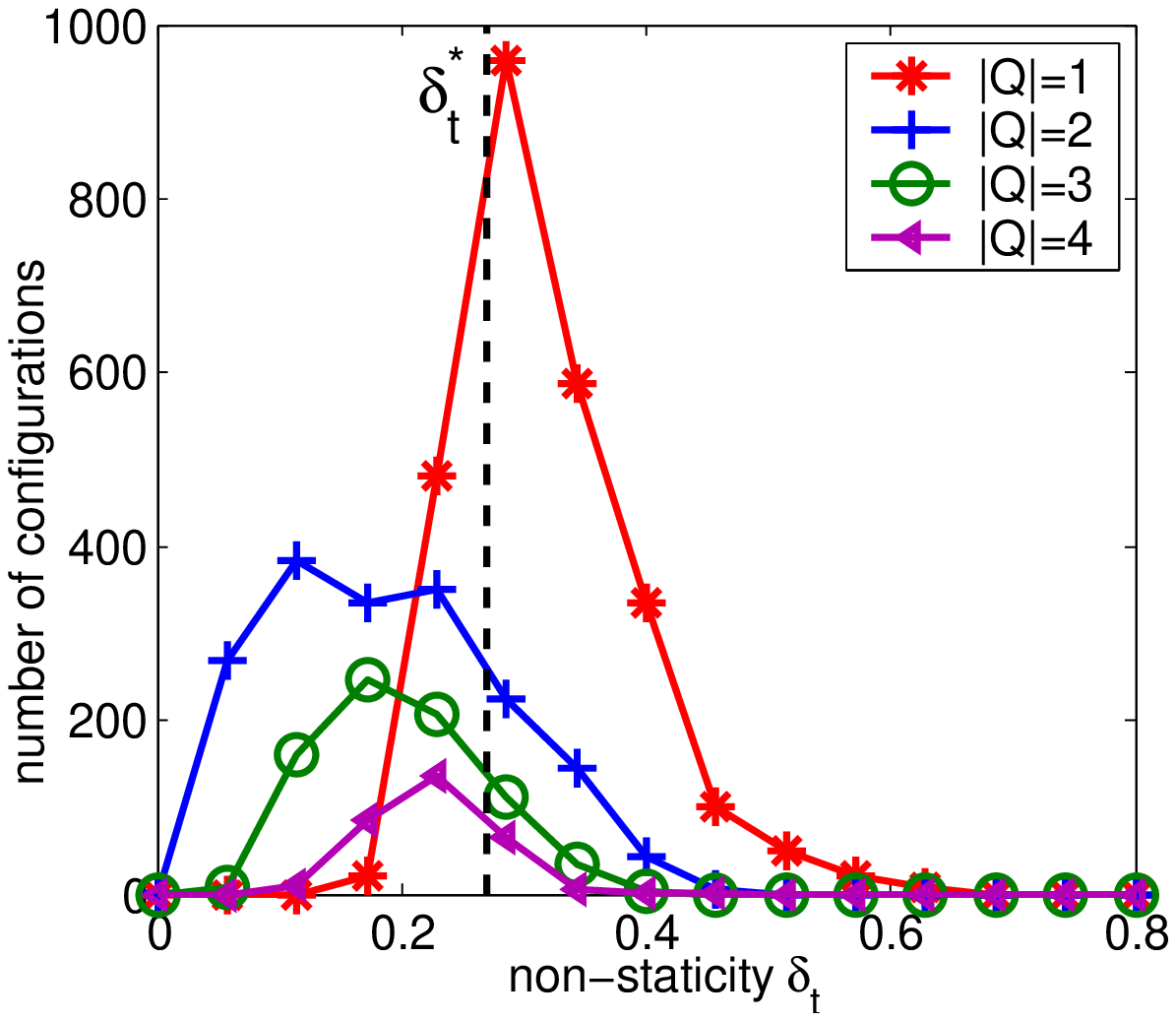}
\includegraphics{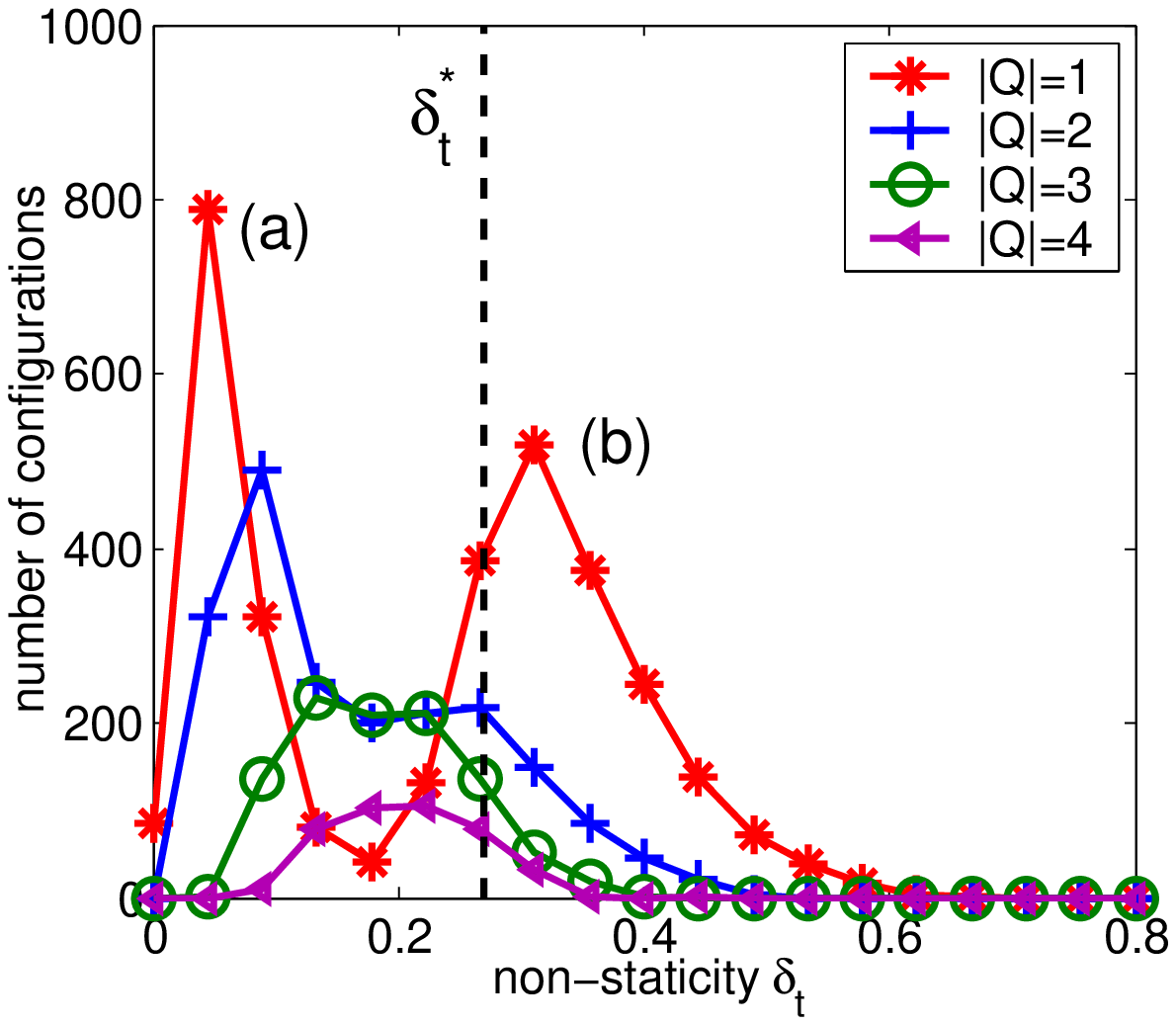}
\includegraphics{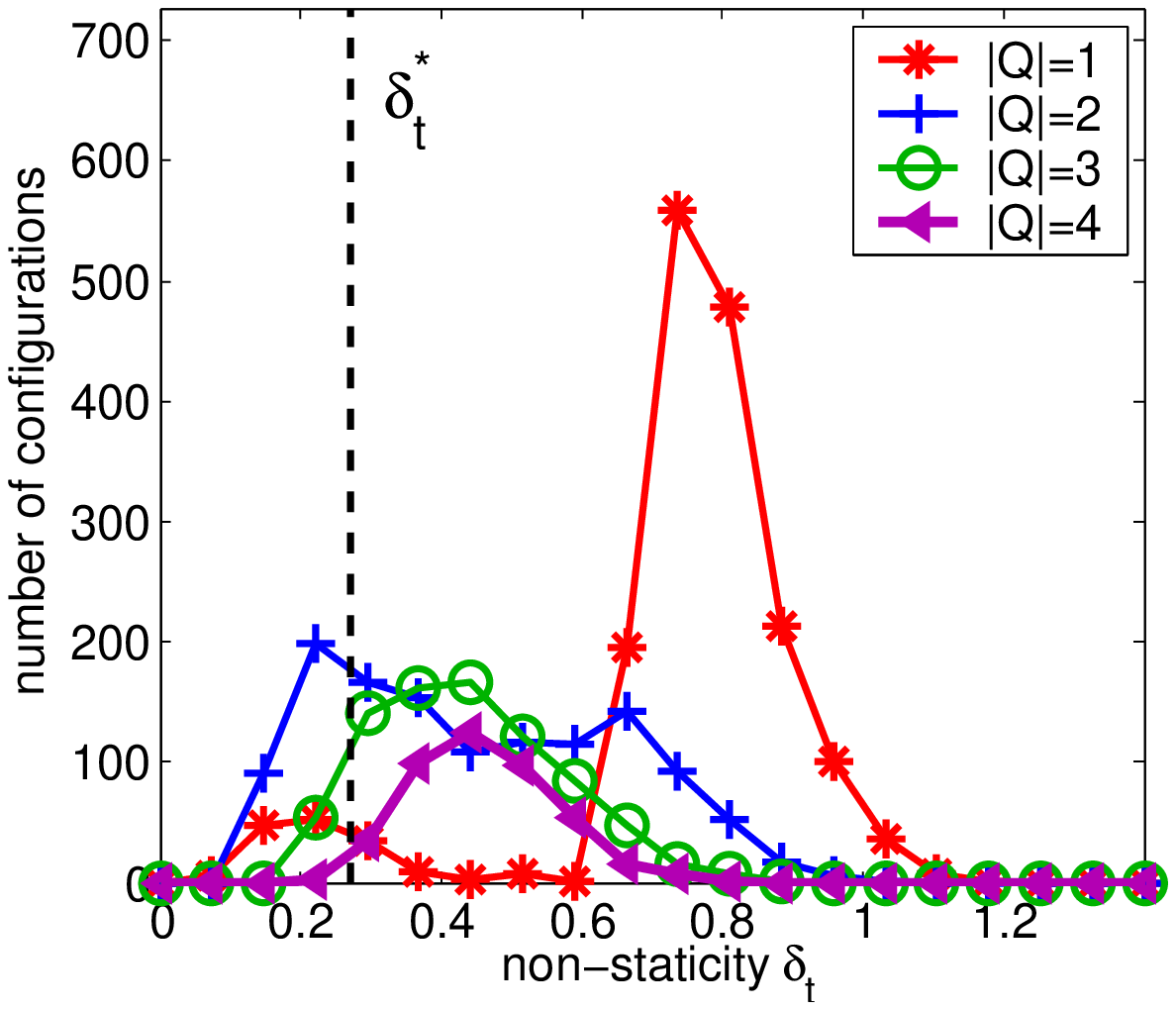}
\includegraphics{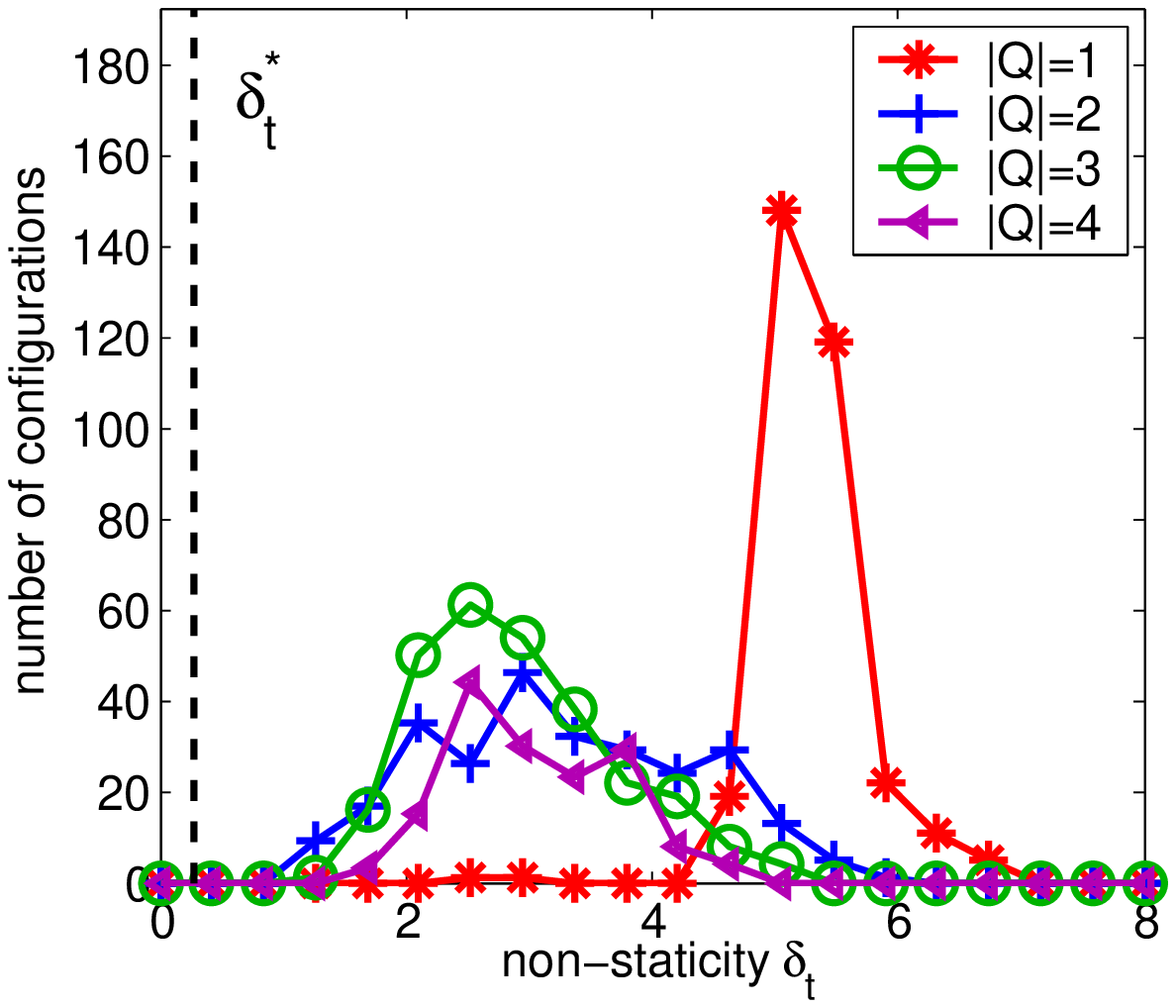}
\caption{Histograms of non-staticity $\delta_t$ for configurations obtained 
when the violation of (anti)self-duality $\delta_F$ is minimal (upper left) 
and when using the extended stopping criterion described in the text (upper 
right), both on a $12^3\! \times\!4$ lattice. The bottom row shows the results 
for $12^3\!\times\!6$ (left) and $12^4$ (right) lattices, also using the 
extended stopping criterion.}\label{fig:nonstaticity_histograms}
\end{figure}

The extended stopping criterion is based on {\em either} the non-staticity 
$\delta_t$ {\em or} the violation of (anti)self-duality $\delta_F$ to pass 
through a minimum. The distribution of (almost) classical solutions on a 
$12^3 \times 4$ lattice with respect to $\delta_t$ is compared for both 
cases in the upper row of Fig.~\ref{fig:nonstaticity_histograms}. 
The vertical line, drawn to guide the eye, represents $\delta_t^{*}$ where the 
bifurcation of 1 into 2 peaks occurs. With the modified stopping criterion, 
which allows for some violation of (anti)self-duality, many perfectly static 
calorons are found which are clearly split into constituents. These are marked 
by ``(a)" on the top right plot in Fig.~\ref{fig:nonstaticity_histograms}, to 
be contrasted with the behavior in the top left plot, and are mainly 
associated to $|Q|=1$ (with some $|Q|=2$) configurations. Yet all satisfy 
$\delta_F<0.2$, and are sufficiently close to a classical solution. In the 
lower row of Fig.~\ref{fig:nonstaticity_histograms} the distribution over 
$\delta_t$ is shown for approximate solutions found on lattices $12^3\times6$ 
and $12^4$ (the symmetric torus), again using the extended stopping criterion. 
This confirms for SU(3) the tendency towards non-dissociated constituents
also found for SU(2)~\cite{PolI} where it is now well understood~\cite{Cool}, 
as was discussed in Sect.~\ref{sec:Cool}.

\begin{figure}[htb]
\vskip3.1cm
\includegraphics{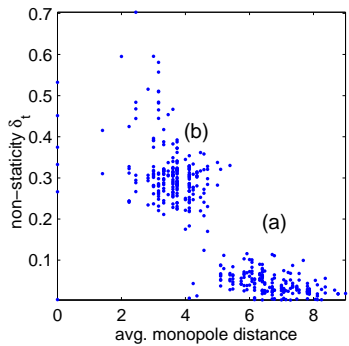}
\caption{Correlation between the non-staticity $\delta_t$ and the average
inter-monopole distance for charge one configurations on $12^3 \times 4$.
\vskip-5mm}\label{fig:monopole_distances}
\end{figure}

As emphasized before, constituent monopoles can be detected through the 
behavior of the Polyakov loop, even when they are close together. This 
is exemplified by Fig.~\ref{fig:monopole_distances} where the non-staticity
is plotted versus the average inter-monopole distance for charge one 
calorons on $12^3\!\times\!4$ lattices. The two clusters marked ``(a)" 
and ``(b)" agree with the configurations associated to the two peaks marked 
in Fig.~\ref{fig:nonstaticity_histograms} for the top right plot, based on 
the extended stopping criterion.

\begin{figure}[!b]
\vskip3.2cm
\includegraphics{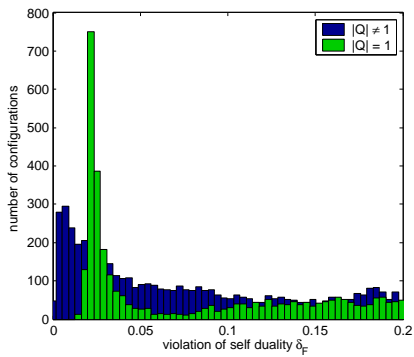}
\caption{Histograms for the violation of (anti) self-duality $\delta_F$,
separately for $|Q|=1$ and $|Q| > 1$ on a lattice of size $12^3 \times 4$.}
\label{fig:violation}
\end{figure}
It is interesting to see how $\delta_F$ is actually distributed. This is shown 
in Fig.~\ref{fig:violation} (where a cut in $\delta_F$ at 0.2 was applied) 
contrasting $|Q|=1$ with $|Q|>1$.  Absence of configurations with charge 
one for $\delta_F\!<\!0.016$ illustrates the non-existence of self-dual 
fields with $|Q|=1$ on the torus.

Regardless of the number of lumps, our method of locating and counting 
monopoles described above is reliable provided the configurations have 
non-trivial holonomy. The number of monopoles is distributed very narrowly 
around $3|Q|$. 
\begin{figure}[htb]
\vskip2.3cm
\includegraphics{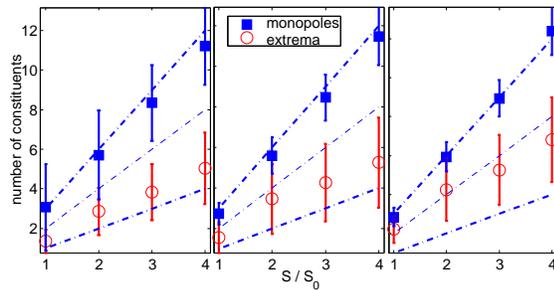}
\caption{Average number of monopoles and of maxima of the topological density
and their standard deviations as function of $|Q|$ for the $12^3 \times 4$ 
lattice, without cuts (left), with a cut $|P_\infty| < 0.2$ (center) and with 
an additional cut $\delta_t < 0.2$ (right).\vskip-7mm}\label{fig:multiplicity}
\end{figure}
This is shown in Fig.~\ref{fig:multiplicity} where the average number of 
monopoles and peaks in the topological charge density are represented as 
a function of $|Q|$, without cuts and with cuts in $|P_{\infty}|$ and 
$\delta_t$. These cuts effect the number of peaks  but not the number of 
monopoles, which however does become more narrowly distributed.  

We have thus seen that the approximate SU(3) caloron solutions produced 
by cooling, even on relatively small lattices, resemble the analytically 
known caloron solutions. The way they manifest themselves depends on the 
aspect ratio of the lattice.

\section{Conclusions}

We have seen that instantons at finite temperature are composed of constituent
monopoles. More details for SU(3) will appear soon~\cite{in_preparation}. The 
hope is of course to be able to develop a calculational scheme to address
questions like monopole condensation in the popular scenario of the dual
superconductor~\cite{Cond}. From the analytic side this is complicated by the 
fact that at low temperature the coupling constant is large, and instantons 
form a dense ensemble. More detailed lattice studies would be welcome. 
A dense ensemble of instantons implies that the monopoles are dense as 
well. The confining electric phase could perhaps be characterized by a 
dual deconfining magnetic phase, where the dual deconfinement is due to 
the large monopole density, similar in spirit to high density induced quark 
deconfinement.

\section*{Acknowledgements}
We thank Margarita Garc\'{\i}a P\'erez, Christof Gattringer and Tony 
Gonz\'alez-Arroyo for discussions. This work was supported in part by 
FOM and by RFBR-DFG (Grant 03-02-04016). FB was supported by FOM and 
E.-M.I. by DFG (Forschergruppe Lattice Hadron Phenomenology, FOR 465).

\end{document}